\documentclass{iopart}
\usepackage{iopams}
\usepackage{amssymb}
\usepackage{amsfonts}
\usepackage{graphicx}
\usepackage{hyperref} 

 \usepackage{color}
\usepackage{url}
\eqnobysec

\newcommand{\text}[1]{\mbox{#1}}
\newcommand{\leb}[1]{\text{d}[#1]}
\newcommand{\dd}[1]{\text{d}#1}
\newcommand{\Det}{\text{det}}
\newcommand{\Pf}{\text{pf}}

\newcommand{\ga}{\gamma}
\newcommand{\Id}{\mathbf{1}}
\newcommand{\IR}{\mathbb{R}}

\newcommand{\U}{{\rm U}}

\newcommand{\nn}{\nonumber}

\newcommand{\diag}{\text{diag}}

\newcommand{\dagg}{^{\dagger}}

\newcommand{\eq}[1]{Eq.~(\ref{#1})}
\newcommand{\noeq}[1]{~(\ref{#1})}
\newcommand{\be}{\begin{equation}}
\newcommand{\ee}{\end{equation}}
\newcommand{\beq}{\begin{eqnarray}}
\newcommand{\eeq}{\end{eqnarray}}

\newenvironment{correction}{ \begin{color}{red}}{\end{color}}
\newenvironment{hasto}{\begin{color}{red}}{\end{color}}

\begin{document}
\title[Smallest eigenvalue distribution in the chGOE with even topology]{The Smallest Eigenvalue Distribution in the Real Wishart-Laguerre Ensemble with Even Topology}
\author{T. Wirtz$^1$, G. Akemann$^2$, T. Guhr$^1$, M. Kieburg$^2$, 
and 
R. Wegner$^2$}
\address{$^1$Fakult\"at f\"ur Physik, Universit\"at Duisburg-Essen, D-47048 Duisburg, Germany\\ 
$^2$Fakult\"at f\"ur Physik, Universit\"at Bielefeld, D-33501 Bielefeld, Germany
}

\ead{tim.wirtz@uni-due.de}

\begin{abstract}
We consider rectangular random matrices of size $p\times n$ belonging to the real Wishart-Laguerre  ensemble also known as the chiral Gaussian orthogonal ensemble. This ensemble appears in many applications like QCD, mesoscopic physics, and time series analysis. We are particularly interested in the distribution of the smallest non-zero eigenvalue and the gap probability to find no eigenvalue in an interval $[0,t]$. While for odd topology $\nu=n-p$ explicit closed results are known for finite and infinite matrix size, for even $\nu>2$ only recursive expressions in $p$ are available. The smallest eigenvalue distribution as well as the gap probability for general even $\nu$ is equivalent
to expectation values of characteristic polynomials raised to a half-integer power. The computation of such averages is done via a combination of skew-orthogonal polynomials and bosonisation methods. The results are given in terms of Pfaffian determinants both at finite $p$ and in the hard edge scaling limit ($p\to\infty$ and $\nu$ fixed) for an arbitrary even topology $\nu$. Numerical simulations for the correlated Wishart ensemble illustrate the universality of our results 
in this particular limit. These simulations point to a validity of the hard edge scaling limit beyond the invariant case.\\

\noindent{\it keywords: 
real Wishart matrices, gap probability, smallest eigenvalue distribution, 
skew-orthogonal polynomials, bosonisation, hard edge scaling limit,
characteristic polynomials of half integer power
}

\end{abstract}

%\pacs{05.45.Tp, 02.50.-r, 02.20.-a}
%\ams{62H05}

\vfill
% \today

%\submitto{\JPA}

\maketitle

%%%%%%%%%%%%%%%%%%%%%%%%%%%%%%%%%%%%%%%%%%%%%%%%%%%%%%%%
\section{Introduction}

The ensemble of real rectangular $p\times n$ matrices $W$ with independent Gaussian entries is the oldest example of random matrix theory, introduced by Wishart in the context of multivariate statistics~\cite{Wishart}. 
Since then more general ensembles built of complex and quaternionic matrix elements have found a wide area of applications, ranging from physics and mathematics to biology and engineering, e.g.  see Refs.~\cite{Guhr1998189,handbook} for reviews and references. These ensembles got several names and are widely known as Wishart ensembles because of its inventor, Laguerre ensembles because of its relation to the Laguerre polynomials, and chiral Gaussian orthogonal, unitary, or symplectic  ensembles hinting to their transformation properties. They are concerned with the singular value statistics of $W$ while in the case of the complex eigenvalue statistics the name Ginibre ensemble is more common.
  
Despite the fact that the real ensembles are more versatile than their complex and especially their quaternion counterpart, those ensembles are at the same time technically challenging. This is particularly true when correlations among the matrix elements are introduced. The introduction of correlations can be done in three ways, either by a correlation matrix  resulting in the correlated Wishart ensemble as it is the case in the analysis of real time series \cite{LCBP99,muirhead,STRF08,RKG10,RAP10,TEP11,RKGZ12,S12,VZ12,Vin,WG,VS14,VB14,WWG14,WKG}, by extending to non-Gaussian probability distributions of the matrix elements \cite{BJJNPZ02,BJN03,BJNPZ07} or by adding constraints \cite{chen,AV,KVZ}. 

When analysing the spectral statistics of the positive definite matrix $WW^T$ one has to distinguish between two kinds of correlations. The first kind involves density correlation functions which can be considered on a global or local scale. The second kind comprises correlations only involving individual eigenvalues such as their distribution or spacing. These correlations are by definition local objects since on a large scale individual eigenvalues with a global separation are usually screened and thus uncorrelated. An efficient tool to compute the second kind of correlations are gap probabilities, meaning that a certain interval is void of eigenvalues.

To emphasize the importance of the distribution of individual eigenvalues we summarise a few applications: The condition number of a matrix $A$ is the ratio of the root of the largest over the smallest non-zero eigenvalues of $AA^T$, which was analysed in a random matrix setting in \cite{Edelman}. The smallest eigenvalues are responsible for chiral symmetry breaking in quantum field theory where real matrices correspond to Quantum Chromodynamics with two colours~\cite{Jac}. Those eigenvalues are very sensitive for fitting lattice QCD data to random matrix results, see \cite{NDW98,WGW98,DN01,AD08,DWW11,LBHW11,AI12,SMN} and in particular \cite{EHKN99,BLP08} for our symmetry class, for the discussion of the importance of individual eigenvalues in fitting lattice data. In this context, the number of zero-eigenvalues $n-p\equiv \nu\geq0$ corresponds to the gauge field topology,  see \cite{JacTilo} for a review on this topic. 
A further application of these quantities can be found in studying topological insulators, see \cite{carlo} for a recent review. In multivariate statistics the smallest eigenvalue plays an important role in high dimensional inference \cite{Kanasewich,BarnettLewis,Gnanadesikan,LarryWasserman}.

In the uncorrelated Gaussian case all density correlation functions are known most explicitly. 
For finite $p$ and arbitrary $\nu$ the $k$-point correlation functions of the real Wishart-Laguerre ensemble are given by a Pfaffian determinant of a kernel involving skew-orthogonal polynomials \cite{Jac,NF95}. These are expressed in terms of Laguerre polynomials. In the limit $p\to\infty$ the local kernels are universal and are given by the corresponding Bessel-, sine- or Airy-kernel, for the hard-edge, bulk or soft-edge scaling limit, see \cite{For10} for the corresponding expressions and references. We are particularly interested in the Bessel kernel in the microscopic origin limit (hard edge). The universality of this kernel was shown for non-Gaussian ensembles in \cite{Klein} and for the some kinds of correlated Wishart ensembles in \cite{WG,WKG}.

Because the individual, e.g. smallest eigenvalue distribution, can be in principle expressed through a Fredholm Pfaffian \cite{Peter06}  (see also \cite{NDW98}) of the very same kernel this universality is inherited by the individual eigenvalue distributions. 
Due to this fact we can restrict ourselves to the Gaussian case.
Apart from this relation the universality of the smallest eigenvalue 
at the soft edge has been proved explicitly \cite{Soshnikov}. 
Furthermore, it was shown in \cite{AV} that the known expressions for the smallest eigenvalue in the Gaussian ensemble and the one with a fixed trace constraint \cite{chen} agree hinting to stronger universal with respect to non-differentiable deformation.

What is know more explicitly about the distribution of the smallest eigenvalue?
Closed expressions 
in the quadratic case $\nu=0$ were derived in \cite{Edelman88} (c.f. \cite{Peter93}). 
In \cite{Edelman} an exact recursive scheme in the matrix size $p$ was set up, leading to closed results for arbitrary $p$ for $\nu=0,1,2,3$ only. 
A distinct structure including both polynomials and Tricomi's confluent  hypergeometric functions was observed to hold for $\nu$ even and odd, respectively. 
Exact results for finite $p$ and in the hard-edge scaling limit were derived in \cite{NF98} for all odd $\nu$ fixed and extended to the $k$-th smallest eigenvalue in \cite{DN01}. Both results are represented as a Pfaffian determinant, a structure that was unnoticed in the recursion of \cite{Edelman}.
Simple approximations following the idea of a Wigner surmise were tested in \cite{ABPS}. An efficient numerical algorithm was used to directly compute the Fredholm Pfaffian expression in \cite{SMN}, and most recently
an extension to the correlated Wishart case was presented in \cite{WG}. 

The goal of the present article is the derivation of a Pfaffian representation for the case of even $\nu$ at finite and infinite $p$ announced in \cite{AGKWWprl}. Thus we aim at completing the picture of the smallest eigenvalue distribution in the real Wishart-Laguerre ensemble. 
The idea in \cite{NF98,DN01} that we will use is to represent the smallest eigenvalue and the gap probability as an expectation value of powers of characteristic polynomials. For odd $\nu$ half integer powers appear, which is the technical problem we have to solve. The tool we apply are skew-orthogonal polynomials with a non-standard weight function containing a square root. The determination of these polynomials is then based on the method of Grassmann variables and bosonisation~\cite{Som07,LSZ08,KSG09}, a particular case of the supersymmetry method, see \cite{Efetovbook,Guhrbook} and references therein.

The relevance of such expectation values of characteristic polynomials including half integer powers has been advocated independently and solved in a few special cases in \cite{AY}, motivated mainly from applications to Quantum Chaos. There, the Gaussian orthogonal ensemble is considered. We present results for the chiral Gaussian orthogonal ensemble (chGOE) for  characteristic polynomials raised to an arbitrary half-integer power. 

The outline of the work is as follows: In section \ref{sec:Problem}, we formulate the problem in terms of expectation values of characteristic polynomials. These can be computed 
by introducing non-standard polynomials that are skew-orthogonal with respect to a weight function containing a square root, see section \ref{sec:Pfaffianstructure}. When expressed in terms of these polynomials both the gap probability and the smallest eigenvalue distribution exhibit a Pfaffian structure. 
The building blocks appearing in these expressions, the partition function, the polynomials and their kernel are computed in section \ref{sec:finite-p-results}. Here we also summarise our exact results for arbitrary $p$ and odd $\nu$. In section \ref{sec:AsymptoticExpressions} we take the microscopic origin limit ($p\to\infty$ and $\nu$ fixed and even). Furthermore, we illustrate our findings by numerical simulations, including the correlated Wishart case which follows the same universal predictions, see section~\ref{sec:numericalsimulations}. Our conclusions and discussion of open problems are presented in section \ref{sec:conclusion}.

%%%%%%%%%%%%%%%%%%%%%%%%%%%%%%%%%%%%%%%%%%%%%%%%%%%%%%%%%%%%%%%%%
\section{Formulation of the Problem}\label{sec:Problem}

We consider  the singular value statistics close to the origin of  the real Wishart-Laguerre ensemble. To this end, we take $W$ to be a rectangular matrix of size $p\times n$ with ``rectangularity'' $\nu=n-p\geq0$ and real entries $W_{ij}\in \IR$. In QCD the ``rectangularity'' $\nu$ is identified with the index of the Dirac-operator and, thus, with the topological charge of the gauge field configuration, c.f. \cite{JacTilo}. The entries of $W$ are drawn from a Gaussian distribution with row-wise correlations~\cite{muirhead}, 
\begin{equation}
\label{eq:intro:wishartP} 
P(W|C) \sim \exp\left(-\frac{1}{2}\tr WW^T C^{-1}\right).
\end{equation}
Sometimes also doubly correlated Wishart random matrices are considered to model spatio-temporal correlations, e.g. see~\cite{RAP10,TEP11,S12,WWG14}.
The singular value statistics of $W$ are completely determined by the eigenvalue statistics of the matrix $WW\dagg$ which  is called the Wishart  correlation matrix. The measure on the space of rectangular matrices $\leb{W}$ is the flat measure, the product of all independent differentials.
In all of our analytic computations we take $C=\Id_p$, i.e. we consider the uncorrelated Wishart model. 
Only in section~\ref{sec:numericalsimulations} we argue that generically also the smallest eigenvalue of the correlated model follows the universal distribution derived in section~\ref{sec:AsymptoticExpressions}.

In order to consider the statistics of the  eigenvalues of the Wishart matrix 
we diagonalize $WW^T =OX O^T$, where $X=\diag(x_1,\dots,x_p)>0$ and $O\in \text{O}(p)$ is an orthogonal matrix.
This leads to the following normalised
joint probability distribution function (jpdf) of the eigenvalues, 
 e.g. see \cite{Mehta,For10}
\begin{equation}
P(X)\equiv\frac{1}{Z_{p,\nu}}\left|\Delta_p(X)\right|\prod_{i=1}^{p}x_i^{(\nu-1)/2}\exp\left(-x_i/2\right)\label{eq:intro:jpdf}~.
\end{equation}
The constant $Z_{p,\nu}$ is  the partition function and, hence, the inverse of the normalisation constant. It is a Selberg integral \cite{Mehta} and explicitly reads
\begin{eqnarray}
 \fl Z_{p,\nu}\equiv
\prod_{i=1}^p\int\limits_0^\infty\dd{x_i}
x_i^{(\nu-1)/2}e^{-x_i/2}
\left|\Delta_p(X)\right|=2^{p(p+\nu)/2} \prod_{j=0}^{p-1}\frac{\Gamma\left[(j+3)/2\right]\Gamma\left[(j+\nu+1)/2\right]}{\Gamma\left[3/2\right]}\nonumber\\
\label{eq:GapNormalization}
\end{eqnarray}
 for arbitrary $\nu\geq-1$. The term $\Delta_p(X)=\prod_{i>j}^p(x_i-x_j)$ denotes the Vandermonde determinant.
The variables $x_i$ coincide with the squares of the singular values of $W$
and typically describe the  low lying eigenvalues $\pm i\sqrt{x_i}$ of the QCD-Dirac operator
\cite{Jac,JacTilo}.

For the smallest non-zero eigenvalues two kinds of large-$p$ limits have to be distinguished. If we take $n$ and $p$ to infinity while $c\equiv p/n$, $c\in(0,1]$, is kept fixed, the macroscopic density of Marchenko-Pastur detaches from the origin. The level density vanishes with a square root behaviour at both endpoints which are called soft edge. In this situation both, the largest \cite{Peter93,TW94} and the smallest eigenvalue, are Tracy-Widom distributed \cite{Soshnikov}.

In contrast to this soft edge scaling we can fix the index
$\nu=n-p$ when taking $n$ and $p$ to infinity. Then the macroscopic density behaves as an inverse square root at the origin, also known as the hard edge. The corresponding scaling is called microscopic origin limit. This limit will be considered in section~\ref{sec:AsymptoticExpressions}.
The microscopic level density \cite{Jac,JacTilo} and all $k$-point correlation functions are given by the Bessel-kernel \cite{NF95} which is universal \cite{Klein}.
In principle all individual eigenvalue distributions including the smallest follow from these density correlations for arbitrary fixed $\nu$. They are given by the Fredholm Pfaffian of the Bessel kernel~\cite{Peter06}. 
In particular the limiting distribution of the smallest eigenvalue, $P_{\nu}(x)$, follows the microscopic spectral density, $\rho_\nu(x)$, for small $x$, in particular $P_{\nu}(x)\approx\rho_\nu(x)$. The level density of the matrix $WW^\dagger$ is~\cite{FNH}
\begin{equation}
 \label{eq:intro:density}
\fl \rho_\nu(u)=\frac{1}{4}\left( J_\nu(\sqrt{u})^2-J_{\nu-1}(\sqrt{u})J_{\nu+1}(\sqrt{u})\right)
+\frac{1}{4\sqrt{u}} J_\nu(\sqrt{u})\left(1-\int\limits_0^{\sqrt{u}}dsJ_\nu(s) \right)
\end{equation}
with $J_\nu$ the Bessel function of the first kind.

At finite $p$ and $\nu$ the distribution of the smallest eigenvalue, $P_{p,\nu}(t)$,  can be derived via
the gap probability $E_{p,\nu}(t)$. This gap probability is the probability to find no eigenvalue in the interval $[0,t]$. 
Starting from the jpdf~\eref{eq:intro:jpdf} we immediately have 
\begin{eqnarray}
E_{p,\nu}(t) &=&\frac{1}{Z_{p,\nu}}\prod_{i=1}^p\int\limits_t^\infty\dd{x_i}
x_i^{(\nu-1)/2}e^{-x_i/2}
\left|\Delta_p(X)\right| 
\label{eq:definitionE}\\
 &=&  \frac{1}{Z_{p,\nu}}e^{-pt/2}\prod_{i=1}^p\int\limits_0^\infty\dd{x_i}
(x_i+t)^{(\nu-1)/2}e^{-x_i/2}
\left|\Delta_p(X)\right| .\label{eq:EShifted}
\end{eqnarray}
In the second line we have shifted $x_i\to x_i+t$ for all $i=1,\ldots,p$ leaving the Vandermonde determinant invariant.

Let us define the expectation value for any integrable quantity $f(X)$ that only depends on the eigenvalues,
\be
\langle f(X)\rangle_{p,\nu}\equiv \frac{1}{Z_{p,\nu}}\prod_{i=1}^p\int\limits_0^\infty\dd{x_i}
x_i^{(\nu-1)/2}e^{-x_i/2}\, f(X)
\left|\Delta_p(X)\right|.
\label{vevdef}
\ee
Then the gap probability~\eref{eq:EShifted} can be written as an expectation value of a characteristic polynomial to a certain power with respect to a different partition function with fixed index $\nu=1$,
\be
E_{p,\nu}(t)=e^{-pt/2}\frac{Z_{p,1}}{Z_{p,\nu}}
\left\langle \Det^{(\nu-1)/2}\left(X+t\Id_p\right)\right\rangle_{p,1}.
\label{gapvev}
\ee
For $\nu=2k+1$ odd the determinant is raised to an integer power $\gamma\equiv(\nu-1)/2=2k$. Such expectation values have been computed for finite dimension \cite{NF98} as well as in the microscopic origin limit \cite{DN01}. For $\nu=2k$ even the expectation value is taken of a half integer power of a determinant. This average was up to now an open problem and is computed in the present work. 
The special cases with $\nu=0$ \cite{Peter93} and $\nu=2$ \cite{AV} were computed in the microscopic limit using a different route, where the latter result is based on the recursive construction of Ref.~\cite{Edelman} that only yields closed form expressions for finite $p$ at $\nu=0$ and $\nu=2$ even.

Next we turn to the distribution of the smallest eigenvalue.
This distribution follows from the gap probability by setting the smallest eigenvalue equal to $t$, say $x_1=t$. Hence it is the first derivative in $t$ of the gap probability~\eref{eq:definitionE},
\begin{eqnarray}
\label{eq:gapdistpmin}
 P_{p,\nu}(t) = -\frac{\text{d}}{\text{d}t}E_{p,\nu}(t)
\quad \Leftrightarrow \quad  
E_{p,\nu}(t) = 1- \int\limits_0^t \dd{t'} P_{p,\nu}(t'),
\end{eqnarray}
where the second  relation fixes the normalisation. Rather than first computing 
$E_{p,\nu}(t)$ and then differentiating with respect to $t$ we can directly start with $P_{p,\nu}(t)$,
\begin{eqnarray}
 \fl P_{p,\nu}(t) &=&  \frac{p}{Z_{p,\nu}}\,
t^{(\nu-1)/2}e^{-t/2}
\prod_{i=2}^p\int\limits_t^\infty\dd{x_i}
x_i^{(\nu-1)/2}e^{-x_i/2}(x_i-t)
|\Delta_{p-1}(x_{j\geq2})|\label{eq:definitionP}\\
\fl&=&  \frac{p}{Z_{p,\nu}}\,
t^{(\nu-1)/2}e^{-pt/2}\prod_{i=2}^p\int\limits_0^\infty\dd{x_i}
x_i(x_i+t)^{(\nu-1)/2}e^{-x_i/2}
|\Delta_{p-1}(x_{j\geq2})|,\label{eq:PShifted}
\end{eqnarray}
where $\Delta_{p-1}(x_{j\geq2})$ does not contain $x_1$.
We recall that the smallest eigenvalue is chosen as $x_1=t$. Therefore the absolute value of these terms can be dropped. In the second line we have again 
shifted $x_i\to x_i+t$ for all $i=2,\ldots,p$. Consequently also 
$P_{p,\nu}(t)$ can be written as an expectation value,
\be
\fl P_{p,\nu}(t)=p\,t^{(\nu-1)/2}e^{-pt/2}\frac{Z_{p-1,3}}{Z_{p,\nu}}
\left\langle \Det^{(\nu-1)/2}\left(X+t\Id_{p-1}\right)\right\rangle_{p-1,3}.
\label{1stvev}
\ee
Here the expectation value is with respect to a partition function of $p-1$ eigenvalues with $\nu=3$ fixed, accounting for the extra factor $x_i$ in \eq{eq:PShifted}.

Once again for $\nu=2k+1$ the expectation value of determinants to an integer power $k$ is known. Our task reduces to the computation  the half integer case $\nu=2k$. Summarising the problem we consider the following type of expectation values
\be
\left\langle 
\Det^{-1/2}\left(X+t\Id_p\right)
\prod_{l=1}^k\Det\left(X+t_l\Id_p\right)
\right\rangle_{p,\nu}.
\label{ourvevs}
\ee
Such problems were advocated independently in Ref.~\cite{AY} for the Gaussian orthogonal ensemble also including more than one square root in the denominator. While special cases have been computed in Ref.~\cite{AY}, having different applications in mind, we determine Eq.~\eref{ourvevs}, leading us to the gap probability and the smallest eigenvalue distribution at finite matrix dimension and in the microscopic limit.

As a final remark the gap probability\noeq{eq:definitionE} has been studied in more detail for even rectangularity $\nu=2k$ in \cite{WG}, where a dual supermatrix model was found. Although the supermatrix model is invariant under the action of a particular supergroup, it was not solved in \cite{WG}, because of non-trivial subtleties related to the necessary diagonalisation. We circumvent these subtleties by combining the method of Grassmann variables with the theory of orthogonal polynomials.

%%%%%%%%%%%%%%%%%%%%%%%%%%%%%%%%%%%%%%%%%%%%%%%%%%%%%%%%%
\section{Pfaffian Structure and Non-Standard Skew-Orthogonal Polynomials}\label{sec:Pfaffianstructure}

We tackle the expectation value~\eref{ourvevs} by including the unwanted inverse half-integer power of the characteristic polynomial into the weight function. The remaining integer powers can be expressed in terms of skew orthogonal polynomials (SOP) using standard techniques. The difficulty is thus shifted into finding the SOP with respect to the $t$-dependent weight 
\begin{eqnarray}
  \label{eq:weightfunction}
 w_\ga(x,t) = \frac{x^\ga}{\sqrt{x+t}}e^{-\eta x/2}.
\end{eqnarray}
The exponent $\ga$ is equal to $\ga=0$ for the gap probability and to $\ga=1$ for the smallest eigenvalue. Thus we calculate the SOP in a unifying way for both quantities. The auxiliary parameter $\eta$ is set to unity unless otherwise stated. It is needed to generate the polynomials of an odd order.

We seek monic, parameter dependent polynomials $R_{j}^{(\ga)}(y,t)=y^j+\ldots$ that are skew-orthogonal with respect to the anti-symmetric product
\begin{eqnarray}
\left<f,g\right>_{t}& =& \int\limits^\infty_0\dd{y}\int\limits_0^\infty\dd{x} \frac{y-x}{|y-x|}w_\ga(x,t)w_\ga(y,t)f(x)g(y)\nonumber\\
&=&\int\limits^\infty_0\dd{y}\int\limits_0^y\dd{x} w_\ga(x,t)w_\ga(y,t)\left(f(x)g(y)-f(y)g(x)\right)\label{eq:scalarproduct}
\end{eqnarray}
for two arbitrary  integrable functions $f$ and $g$. In particular the polynomials have to fulfill the relations
\begin{eqnarray}
\label{eq:definitionorhtogonalpoly}
\fl\quad\quad\left<R_{2j+1}^{(\ga)},R_{2i}^{(\ga)}\right>_{t} = r_j^{(\ga)}(t)\delta_{ij}~, \quad  \left<R_{2j}^{(\ga)},R_{2i}^{(\ga)}\right>_{t} =  \left<R_{2j+1}^{(\ga)},R_{2i+1}^{(\ga)}\right>_{t}=0
\end{eqnarray}
for $p=2L$ even.
The parameter dependent constants $r_{i}^{(\ga)}(t)$ are their normalization constants. In the case $p=2L+1$ odd the polynomials have to satisfy an additional condition \cite{Mehta}
\begin{eqnarray}
\fl\quad&\quad\quad\left<\widehat{R}_{2j+1}^{(\ga)},\widehat{R}_{2i}^{(\ga)}\right>_{t} = r_j^{(\ga)}(t)\delta_{ij}~, \quad  \left<\widehat{R}_{2j}^{(\ga)},\widehat{R}_{2i}^{(\ga)}\right>_{t} =  \left<\widehat{R}_{2j+1}^{(\ga)},\widehat{R}_{2i+1}^{(\ga)}\right>_{t}=0,&\nonumber\\
 \fl &\quad\quad\int\limits_0^\infty\dd{x}\widehat{R}_{i}^{(\ga)}(x,t) w_\ga(t;x) = \delta_{i,2K},& \label{eq:additionalrequiermentpodd}
\end{eqnarray}
which fixes the normalization of the polynomial of highest order $\widehat{R}_{2K}^{(\ga)}(x,t)$ with $K\geq L$. Note that the normalization constants $r_j^{(\ga)}(t)$ are the same as for the case $p=2L$. The reason is the following relation between the two different kinds of the polynomials
\begin{eqnarray}
\widehat{R}_{j}^{(\ga)}(y,t)=R_{j}^{(\ga)}(y,t)-\frac{\int_0^\infty \dd x w_\ga(x,t) R_{j}^{(\ga)}(x,t)}{\int_0^\infty \dd x w_\ga(x,t) R_{2K}^{(\ga)}(x,t)}R_{2K}^{(\ga)}(y,t) \label{polynomial-rel}
\end{eqnarray}
for $j<2K$ and $\widehat{R}_{2K}^{(\ga)}(y,t)=R_{2K}^{(\ga)}(y,t)/\int_0^\infty \dd x w_\ga(x,t) R_{2K}^{(\ga)}(x,t)$. Therefore we concentrate on the polynomials $R_{j}^{(\ga)}(y,t)$ only. We emphasize that the additional condition $\int_0^\infty\dd{x}\widehat{R}_{i}^{(\ga)}(x,t) w_\ga(t;x) = \delta_{i,2K}$ can be also replaced by other conditions. For example in the framework of \cite{Kie12} the condition would read $\int_0^\infty\dd{x}\widehat{R}_{i}^{(\ga)}(x,t) w_\ga(t;x) = \delta_{i,0}$ for all $i>0$ which has other advantages in the calculation.

We define a new partition function
\beq
\fl\quad\quad\quad Z_{p,\ga}(t)&\equiv&\prod_{i=1}^p\int\limits_0^\infty\dd{x_i}
w_\ga(x_i;t)
\left|\Delta_p(X)\right|\nonumber\\
\fl\quad\quad\quad\quad&=& p! \prod_{j=0}^{\lfloor p/2\rfloor-1}r_j^{(\ga)}(t)\left\{\begin{array}{cl} \displaystyle\int_0^\infty \dd x w_\ga(x,t) R_{2L}^{(\ga)}(x,t), & p=2L+1, \\ 1, & p=2L \end{array}\right.\nonumber\\
\fl&=&Z_{p,2\ga+1}\left\langle \Det^{-1/2}(X+t\Id_p)\right\rangle_{p,2\ga+1},\label{part-norm}
\eeq
where the second line is a general relation between the normalization constants of  the SOP and the partition function~\cite{Mehta}. The floor function $\lfloor p/2\rfloor$ yields the largest integer smaller than or equal to $p/2$. Employing this partition function we introduce  a new parameter dependent expectation value of an integrable observable $f$,
\be
\langle f(X)\rangle_{p,\ga}^t\equiv \frac{1}{Z_{p,\ga}(t)}\prod_{i=1}^p\int\limits_0^\infty\dd{x_i}
w_\ga(x_i;t)\, f(X)
\left|\Delta_p(X)\right|.
\label{vevdef-t}
\ee
The parametric dependence on $t$ is indicated through the superscript.

In this framework the gap probability~\eref{gapvev} and the distribution of the smallest eigenvalue~\eref{1stvev} (both for $\nu=2k$) read
\beq
\fl \quad\quad\quad\quad E_{p,2k}(t)&=&e^{-pt/2}\frac{Z_{p,0}(t)}{Z_{p,2k}}
\left\langle \Det^{k}\left(X+t\Id_p\right)\right\rangle_{p,0}^t,
\label{gapvevt}\\\quad\quad\quad\quad
\fl P_{p,2k}(t)&=&p\,t^{(2k-1)/2}e^{-pt/2}\frac{Z_{p-1,1}(t)}{Z_{p,2k}}
\left\langle \Det^{k}\left(X+t\Id_{p-1}\right)\right\rangle_{p-1,1}^t.
\label{1stvev.b}
\eeq
It is worth emphasizing that now only integer powers appear in both expressions. Thus, we can apply the results from the
literature for general weight functions, c.f. \cite{NF98,BS,KG}.
For $k=2m$ even, we obtain the following Pfaffian expression with a $2m\times 2m$ dimensional kernel, $\mathcal K_{p+k}(\kappa_a,\kappa_b)$, 
\begin{equation}
\fl\quad\left\langle \prod_{a=1}^k\Det(X-\kappa_a\Id_p)\right\rangle_{p,\ga}^t=
\frac{p!Z_{p+k,\ga}(t)}{(p+k)!Z_{p,\ga}(t)}\frac{1}{\Delta_{k}(\kappa)}
\Pf_{1\leq a,b \leq k}\left[\mathcal K_{p+k}(\kappa_a,\kappa_b,t) \right],\label{eq:partitionfunctionkeven}
\end{equation}
while for $k=2m+1$ we have 
\begin{equation}
\fl\quad\eqalign{\left\langle \prod_{a=1}^k\Det(X-\kappa_a\Id_p)\right\rangle_{p,\ga}^t&=\frac{p!Z_{p+k+1,\ga}(t)}{(p+k+1)!Z_{p,\ga}(t)}
\frac{1}{\Delta_{k}(\kappa)} \\
&\times
\Pf_{1\leq a,b \leq k}\left[
\begin{array}{ll}
\mathcal K_{p+k+1}(\kappa_a,\kappa_b) & \mathcal F_{p+k+1}(\kappa_a,t)\\
-\mathcal F_{p+k+1}(\kappa_a,t)&0\\
\end{array}
\right]. }\label{eq:partitionfunctionkodd}
\end{equation}
The kernels inside the Pfaffians are given by
\begin{eqnarray}
\fl\mathcal K_{l}(\kappa_a,\kappa_b,t) =\left\{\begin{array}{cl} \displaystyle\sum_{j=0}^{(l-2)/2} \frac{R_{2j+1}^{(\ga)}\left(\kappa_a,t\right)R_{2j}^{(\ga)}\left(\kappa_b,t\right)
-R_{2j+1}^{(\ga)}\left(\kappa_b,t\right)
R_{2j}^{(\ga)}\left(\kappa_a,t\right)}{r_j^{(\ga)}(t)}, & \\ &\hspace*{-1cm} l\in2\mathbb{N}, \\
\displaystyle \sum_{j=0}^{(l-3)/2} \frac{\widehat{R}_{2j+1}^{(\ga)}\left(\kappa_a,t\right)\widehat{R}_{2j}^{(\ga)}\left(\kappa_b,t\right)
-\widehat{R}_{2j+1}^{(\ga)}\left(\kappa_b,t\right)
\widehat{R}_{2j}^{(\ga)}\left(\kappa_a,t\right)}{r_j^{(\ga)}(t)}, & \\ &\hspace*{-1cm} l\in2\mathbb{N}+1. \end{array}\right.\nonumber\\
\fl
\label{eq:Pfaffiankernel}
\end{eqnarray}
The case of $k=2m+1$ odd is obtained here from the case $k=2m+2$ even by introducing an additional determinant in the average depending on the dummy variable $\kappa_{2m+2}$. This variable is sent to infinity such that the additional row and column in Eq.~\eref{eq:partitionfunctionkodd} reads
\begin{equation}
\mathcal F_{l}(\kappa_a,t)=-\lim\limits_{\kappa_{2m+2}\to\infty}\frac{\mathcal K_{l}(\kappa_a,\kappa_{2m+2},t)}{\kappa_{2m+2}^{l-1}}.
\end{equation}
This limit is independent of $l$ being even or odd.

Once we have determined the SOP and their kernel we have to take the limit, $\kappa_a\to -t$ for all $\kappa_a$. This yields derivatives of the polynomials $R_{i}^{(\ga)}\left(\kappa_b,t\right)$ because of l'Hospital's rule. Moreover we could also just take part of the $\kappa_a\to-t$ which is needed to calculate the distributions of the second smallest eigenvalue, the third smallest eigenvalue etc., see Ref.~\cite{DN01}. The distributions of the smallest eigenvalue for QCD with dynamical quarks can be found in this way, too, cf. Refs.~\cite{DN01}.

In the next section we explicitly compute the SOP. For this computation it is helpful to understand also these polynomials as expectation values
\cite{Eynard,GP02},
\beq
R_{2j}^{(\ga)}(y,t) &=&\left\langle \Det\left(y\Id_{2j}-X\right)\right\rangle_{2j,\ga}^t,
\label{eq:polynomialsevendegree}\\
 R_{2j}^{(\ga)}(y,t) &=& \left\langle \Det\left(y\Id_{2j+1}-X\right)(y+c_j(t)+\Tr X)\right\rangle_{2j+1,\ga}^t\nn\\
 &=&\left.\left(y+c'_j(t)-2 \frac{\partial}{\partial \eta} \right)R_{2j}^{(\ga)}(y,t)\right|_{\eta=1}\nn\\
 &=&\left(y+\widehat{c}_j(t)-2 y\frac{\partial}{\partial y}-2 t\frac{\partial}{\partial t} \right)R_{2j}^{(\ga)}(y,t).\label{eq:polynomialsodddegree}
\eeq
These two last relations also hold in a much more general framework where $w_\ga(x,t)w_\ga(y,t)(y-x)/|y-x|$ is replaced by an arbitrary anti-symmetric two-point weight $g(x,y)=-g(y,x)$ \cite{AKP10}.

The odd polynomials are not unique \cite{Eynard,Mehta} which is reflected in an ambiguous constant $c_j(t)$. This constant can depend on $t$ and the index $j$ but is independent of $y$. This dependence is the reason why one could absorb the derivative of the normalization into $c'_j(t)$ and rephrase the derivative in $\eta$ as a derivative in $t$ and $y$, yielding a new constant in $\widehat{c}_j(t)$.
We will stick to the derivative in $\eta$ here.

The kernel can be directly expressed as an expectation value, too. Rather than computing the individual polynomials~\eref{eq:polynomialsevendegree} and \eref{eq:polynomialsodddegree} and performing the sum~\eref{eq:Pfaffiankernel} one can consider the average 
\begin{eqnarray}
\fl &&\mathcal K_{l}(\kappa_a,\kappa_b,t) = \frac{l(l-1)Z_{l-2,\ga}(t)}{Z_{l,\ga}(t)}(\kappa_a-\kappa_b)\left\langle \Det\left(X-\kappa_a\Id_{l-2}\right)\Det\left(X-\kappa_b\Id_{l-2}\right)\right\rangle_{l-2,\ga}^{t},\nonumber\\
\fl&&\label{kernel-av-1}
\end{eqnarray}
e.g see \cite{APSo} in the hermitian limit or \cite{KG} in the general framework of anti-symmetric two-point weights.
This representation is also useful when proving that the large-$p$ limit for even and odd $p$ yields the same answer. The additional row and column in eq.~\eref{eq:partitionfunctionkodd} is then
\be\label{kernel-av-2}
 \mathcal F_{l}(\kappa_a,t) =\frac{l(l-1)Z_{l-2,\ga}(t)}{Z_{l,\ga}(t)}\left\langle \Det\left(X-\kappa_a\Id_{l-2}\right)\right\rangle_{l-2,\ga}^{t}.
\ee
We emphasize that for $l$ even this function is equal to the polynomial $R_{l-2}^{(\gamma)}(\kappa_a)$, up to a constant.

%%%%%%%%%%%%%%%%%%%%%%%%%%%%%%%%%%%%%%%%%%%%%%%%%%%%%%%%%
\section{Calculation of the Finite $p$ Results}\label{sec:finite-p-results}

We start our calculation by considering the expectation value
\begin{eqnarray}
I_l(\kappa)=\left\langle \prod_{a=1}^k\Det(X-\kappa_a\Id_l)\right\rangle_{l,\ga}^t.\label{average}
\end{eqnarray}
This quantity is a polynomial in the variables $\kappa_a$. The highest power in these variables determines its normalization, $I_l(\kappa)=(-1)^{lk}\kappa_1^l\cdots\kappa_k^l+\ldots$, such that we can omit the normalization constants in the intermediate steps of our calculation. The overall constant can be fixed at the end of the calculation.

In the first step we rewrite Eq.~\eref{average} as an integral over a rectangular real matrix $\widehat{W}$ of dimension $l\times(l+2\ga+1)$,
where in this and the next subsection $2\ga+1$ can be any integer,
\begin{eqnarray}
\fl I_l(\kappa)&\propto& \int\leb{\widehat W}   \prod_{a=1}^k\Det(\kappa_a\Id_l-\widehat{W}\widehat{W}^T) \frac{\exp\left(-\tr \widehat W \widehat W ^T/2\right)}{\sqrt{\Det\left(\widehat W \widehat W ^T+t\Id_{l}\right)}}\nonumber\\
\fl&\propto& \int\leb{\widehat W}   \prod_{a=1}^k\Det(\kappa_a\Id_l-\widehat{W}\widehat{W}^T) \frac{\exp\left(-\tr \widehat W \widehat W ^T/2\right)}{\sqrt{\Det\left( \widehat W ^T\widehat W+t\Id_{l+2\ga+1}\right)}}.\label{calc-1}
\end{eqnarray}
We emphasize that the normalization constants in each of the steps may dependent on $t$. In the second line of Eq.~\eref{calc-1} we replaced $\widehat W \widehat W ^T\to\widehat W ^T\widehat W$ because of the relation between the determinant and the trace, i.e. ${\rm ln}\,\det A=\tr{\rm ln}\,A$,  and the invariance of the trace under circular permutations. The switching of the order of $\widehat W$ and $\widehat W^T$  allows us to avoid the Efetov-Wegner terms appearing in the superspace dual to this average, see for details Refs.~\cite{Efetovbook,RKGZ12,RKG10,KGG09}.

We introduce a real $(l+2\ga+1)$-dimensional vector $v$ to rewrite the single determinant in the denominator as a Gaussian integral. Additionally we express the product of determinants in the numerator as a Gaussian integral  over a rectangular matrix $V$ of dimension $l\times 2k$ whose entries are independent Grassmann variables (anti-commuting variables). For an introduction in supersymmetry we refer to Refs.~\cite{Ber,Efetovbook,Guhrbook} and for the supersymmetry method with general weight to Refs.~\cite{Guhr06,Som07,LSZ08,KGG09,KSG09}. Particularly the bosonisation is described in Refs.~\cite{Som07,LSZ08,KSG09}. The matrix $V$ satisfies the following symmetry under complex conjugation and under Hermitian conjugation,
\begin{equation}\label{conj}
 V^*=V\left[\begin{array}{cc} 0 & \Id_{k} \\ -\Id_{k} & 0 \end{array}\right]\ {\rm and}\ (V^\dagger)^\dagger=-V,\ {\rm respectively}.
\end{equation}
Then, the average reads
\begin{eqnarray}
\fl I_l(\kappa)&\propto&\int\leb{\widehat W}\int\leb{V}\int\leb{v}   \exp\left(-\frac{1}{2}\left[\tr \widehat W \widehat W ^T+\tr \widehat W \widehat W ^T VV^\dagger+ \tr \widehat Wvv^T\widehat W ^T\right]\right)\nonumber\\
\fl&&\times\exp\left(-\frac{1}{2}\left[\tr V^\dagger V\kappa+tv^Tv\right]\right)\label{calc-2}
\end{eqnarray}
with $\kappa=\diag(\kappa_1,\ldots,\kappa_k)\otimes\Id_2$. Because of the symmetry~\eref{conj} the dyadic matrix $VV^\dagger$ behaves like a real symmetric matrix such that we can integrate over the matrix $\widehat{W}$ without symmetrizing the other terms. This integration yields 
\begin{eqnarray}
\fl I_l(\kappa)&\propto& \int\leb{V}\int\leb{v} {\det}^{-1/2}\left(\Id_l\otimes\Id_{l+2\ga+1}+VV^\dagger\otimes\Id_{l+2\ga+1}+ \Id_l\otimes vv^T\right)\nonumber\\
\fl&&\times\exp\left(-\frac{1}{2}\left[\tr V^\dagger V\kappa+tv^Tv\right]\right)\nonumber\\
\fl&\propto& \int\leb{V}\int\leb{v} {\det}^{-1/2}\left([1+v^Tv]\Id_l+VV^\dagger\right)\nonumber\\
\fl&&\times{\det}^{-(l+2\ga)/2}\left(\Id_l+VV^\dagger\right)\exp\left(-\frac{1}{2}\left[\tr V^\dagger V\kappa+tv^Tv\right]\right)\nonumber\\
\fl&\propto&  \int\leb{V}\int\leb{v} [1+v^Tv]^{-(l+2k)/2}{\det}^{1/2}\left([1+v^Tv]\Id_{2k}+V^\dagger V\right)\nonumber\\
\fl&&\times{\det}^{(l+2\ga)/2}\left(\Id_{2k}+V^\dagger V\right)\exp\left(-\frac{1}{2}\left[\tr V^\dagger V\kappa+tv^Tv\right]\right).\label{calc-2.b}
\end{eqnarray}
Again we have used the relation between the determinant and the trace and the invariance of the trace under circular permutations. However we have to remind ourselves that anti-commuting variables are involved such that $\tr (VV^\dagger)^m=-\tr (V^\dagger V)^m$ for any $m\in\mathbb{N}$. This explains the change from negative to positive powers of the determinant in $VV^\dagger$.

In the last step we can choose between two approaches, the generalized Hubbard-Stratonovich transformation~\cite{Guhr06,KGG09} and the superbosonization formula~\cite{Som07,LSZ08}. Both approaches are equivalent~\cite{KSG09}. We choose the superbosonization formula since it directly leads to a compact expression. Since no supermatrices comprising both, bosonic and fermionic, blocks are involved the superbosonization reduces to bosonisation, only. This means that the norm $v^Tv$ is replaced by a positive variable $r$ (the square of the radial part of an ordinary real vector) and the dyadic matrix $V^\dagger V$ is replaced by a self-dual, unitary matrix,
\begin{equation}\label{CSE}
 U=\left[\begin{array}{cc} 0 & -\Id_{k} \\ \Id_{k} & 0 \end{array}\right]U^T\left[\begin{array}{cc} 0 & \Id_{k} \\ -\Id_{k} & 0 \end{array}\right]\in{\rm U}(2k),
\end{equation}
because $V^\dagger V$ is itself self-dual. The set of matrices defined via Eq.~\eref{CSE} is the circular symplectic ensemble first studied by Dyson~\cite{Dyson}. This set is the coset ${\rm CSE}(2k)={\rm U}(2k)/{\rm USp}(2k)$ and has a uniquely induced Haar measure $\dd\mu(U)$ from the unique, normalized Haar measure of the unitary group ${\rm U}(2k)$. In particular 
up to a normalization constant
it is given by $\dd\mu(U)\propto\leb{U}/{\det}^{k-1/2}U$, with $\leb{U}$ the product of differentials of all independent matrix entries of $U$. The superbosonization formula yields
\begin{eqnarray}
\fl I_l(\kappa)&=& C_l^{-1}\int\limits_{{\rm CSE}(2k)} \dd\mu(U) {\det}^{-l/2}U{\det}^{(l+2\ga)/2}\left(\Id_{2k}+U\right) \int_0^\infty\dd r \frac{r^{(l+2\ga-1)/2}}{(1+r)^{(l+2k)/2}}\nonumber\\
\fl&&\times{\det}^{1/2}\left([1+r]\Id_{2k}+U\right)\exp\left(-\frac{1}{2}\left[\tr U\kappa+t\,r\right]\right),\label{calc-3}
\end{eqnarray}
with the normalization constant
\begin{eqnarray}
\fl C_l&=&\int\limits_{{\rm CSE}(2k)} \dd\mu(U) {\det}^{-l/2}Ue^{\tr U/2}\int_0^\infty\dd r \frac{r^{(l+2\ga-1)/2}}{(1+r)^{l/2}}e^{-t\,r/2}.\label{calc-norm}
\end{eqnarray}
The constant follows from the asymptotics for $\kappa\to\infty$.
The powers of the additional terms ${\det}^{-l/2}U$ and $r^{(l+2\ga-1)/2}$ only reflect the nature of the variables from where $U$ and $r$ originate. We underline that the half-integer of the determinants do not cause any problems since the matrices are Kramers degenerate. Therefore the determinants of them are exact squares and the square root is taken such that the result is still a polynomial in the matrix entries.

Starting from expression~\eref{calc-3} we calculate the partition function $Z_{p,\ga}(t)$ in subsection~\ref{subsec:normalization}, the polynomials $R_{j}^{(\ga)}(y,t) $ and the function $\mathcal F_{l}(\kappa_a)$ in subsection~\ref{subsec:polynomials}, and the kernel $\mathcal K_{l}(\kappa_a,\kappa_b)$  in subsection~\ref{subsec:kernel}. In subsection~\ref{subsec:distribution} we collect everything and give explicit expressions for the gap probability and the distribution of the smallest eigenvalue.

%%%%%%%%%%%%%%%%%%%%%%%%%%%%%%%%%%%%%%%%%%%%%%%%%%%%%%%%%%%%%%%%%
\subsection{Normalization Constants}\label{subsec:normalization}

The partition function $Z_{p,\ga}(t)$ is equal to the case $l\to p$ and $k\to 0$ in the integral~\eref{calc-3}. Therefore we have no integral over a circular unitary ensemble and only the integral over $r$ remains, i.e.
\begin{eqnarray}
\fl Z_{p,\ga}(t)=\frac{Z_{p,2\ga+1}}{2^{(p+2\ga+1)/2}\Gamma[(p+2\ga+1)/2]} t^{\gamma+1/2}\int_0^\infty\dd r \frac{r^{(p+2\ga-1)/2}}{(1+r)^{p/2}}e^{-t\,r/2}.\label{norm-1}
\end{eqnarray}
The constant is fixed by the asymptotic behaviour $Z_{p,\ga}(t)=Z_{p,2\ga+1}t^{-p/2}+o(t^{-p/2})$ for $t\to\infty$. The remaining integral is a Tricomi confluent hypergeometric function~\cite{AS-book},
\begin{eqnarray}
 \fl&&{\rm U}(a,b,t)=\frac{1}{\Gamma[a]}\int_0^\infty dz z^{a-1} (1+z)^{b-a-1} e^{-t z},\ {\rm with}\ {\rm Re}\,a,\ {\rm Re}\,t>0\ {\rm and}\ a,b,t\in\mathbb{C}.\nonumber\\
 \fl&&\label{confluent}
\end{eqnarray}
This hypergeometric function was already found in the work by Edelman~\cite{Edelman} and is a crucial ingredient in his recursive formula. The partition function reads
\begin{eqnarray}
\fl Z_{p,\ga}(t)&=&2^{(p-1)(p+2\ga+1)/2}\left(\prod\limits_{j=0}^{p-1}\frac{\Gamma[(j+3)/2]\Gamma[(j+2\gamma+2)/2]}{\Gamma[3/2]}\right)\nonumber\\
\fl&&\times t^{\gamma+1/2} {\rm U}\left(\frac{p+2\ga+1}{2},\frac{2\gamma+3}{2},\frac{t}{2}\right)\label{norm-2}\\
\fl&=&2^{p(p+2\ga)/2}\left(\prod\limits_{j=0}^{p-1}\frac{\Gamma[(j+3)/2]\Gamma[(j+2\gamma+2)/2]}{\Gamma[3/2]}\right){\rm U}\left(\frac{p}{2},\frac{1-2\gamma}{2},\frac{t}{2}\right).\nonumber
\end{eqnarray}
The second equality follows from the Kummer identity of Tricomi's confluent hypergeometric function~\cite{NIST}, ${\rm U}(a,b,t)=t^{1-b}{\rm U}(a+1-b,2-b,t)$. Combining \eref{eq:GapNormalization} and \eref{part-norm}
we have thus obtained the first building block for the Paffian structure,
\begin{eqnarray}
\left\langle \Det^{-1/2}(X+t\Id_p)\right\rangle_{p,\nu}=2^{-p/2}
{\rm U}\left(\frac{p}{2},\frac{2-\nu}{2},\frac{t}{2}\right),
\label{sqrtvev}
\end{eqnarray}
which is even valid for any $\nu\in\mathbb{N}_0$.

In the particular case of the gap probability the constant is
\begin{eqnarray}
 \fl Z_{p,0}(t)=2^{(p-1)/2}\left(\prod\limits_{j=1}^{p}j!\right) t^{1/2} {\rm U}\left(\frac{p+1}{2},\frac{3}{2},\frac{t}{2}\right)=2^{p/2}\left(\prod\limits_{j=1}^{p}j!\right){\rm U}\left(\frac{p}{2},-\frac{1}{2},\frac{t}{2}\right).\nonumber\\
 \fl\label{norm-3}
\end{eqnarray}
We have used the duplication formula of the Gamma function, $\Gamma[z]\Gamma[z+1/2]=2^{1-2z}\sqrt{\pi}\Gamma[2z]$, to simplify the expression.
Also the partition function needed for the distribution of the smallest eigenvalue the partition function takes a simple form,
\begin{eqnarray}
 Z_{p-1,1}(t)&=&2^{(p-4)/2}\left(\prod\limits_{j=1}^{p}j!\right)t^{3/2} {\rm U}\left(\frac{p+2}{2},\frac{5}{2},\frac{t}{2}\right)\nn\\
 &=&2^{(p-1)/2}\left(\prod\limits_{j=1}^{p}j!\right){\rm U}\left(\frac{p-1}{2},-\frac{1}{2},\frac{t}{2}\right).\nonumber\\
 \fl\label{norm-4}
\end{eqnarray}
The second expression of Tricomi's confluent hypergeometric function is employed in Edelman's work~\cite{Edelman}.

The normalizations $r_j^{(\ga)}(t)$ of the SOP can be deduced by combining Eqs.~\eref{part-norm} and \eref{norm-2},
\begin{eqnarray}\label{norm-pol}
 r_j^{(\ga)}(t)&=&\frac{ Z_{2j+2,\ga}(t)}{(2j+2)(2j+1)Z_{2j,\ga}(t)}\\
 &=&2(2j)!\Gamma[2j+2\gamma+2]\frac{{\rm U}\left(j+\gamma+3/2,\gamma+3/2,t/2\right)}{{\rm U}\left(j+\gamma+1/2,\gamma+3/2,t/2\right)}\nonumber\\
 &=&2(2j)!\Gamma[2j+2\gamma+2]\frac{{\rm U}\left(j+1,-\gamma+1/2,t/2\right)}{{\rm U}\left(j,-\gamma+1/2,t/2\right)}.\nonumber
\end{eqnarray}
This term becomes important in the sum in the kernel.

%%%%%%%%%%%%%%%%%%%%%%%%%%%%%%%%%%%%%%%%%%%%%%%%%%%%%%%%%%%%%%%%%
\subsection{Skew-Orthogonal Polynomials}\label{subsec:polynomials}

First we concentrate on the function $\mathcal F_{l}(\kappa_a,t)$, see Eq.~\eref{kernel-av-2}. For this case we set $l\to l-2$, $k\to1$, and $\kappa\to\kappa_a\Id_2$. The $2\times2$ dimensional unitary matrix $U$ only consists of one phase $e^{i\varphi}$ on the diagonal due to its self-duality. Thus we have to calculate the double integral
\begin{eqnarray}
\fl \mathcal F_{l}(\kappa_a,t)&\propto& \int\limits_0^{2\pi} \dd\varphi \frac{\left(1+e^{i \varphi}\right)^{l+2\ga-2}}{e^{i (l-2)\varphi}}e^{\displaystyle -\kappa_a e^{i \varphi}} \int_0^\infty\dd r \frac{r^{(l+2\ga-3)/2}}{(1+r)^{l/2}}e^{\displaystyle-t\,r/2}\left(1+r+e^{i \varphi}\right).\nonumber\\
\fl&&\label{func-1}
\end{eqnarray}
The term $(\eta+r+e^{i \varphi})$ is the only coupling between the two integrals and yields  a sum of two terms, of which each is a product of two functions. The integrals over $r$ are equal to Tricomi confluent hypergeometric functions~\eref{confluent} while the integrals over $\varphi$ are modified Laguerre polynomials in monic normalization,
\begin{equation}\label{Laguerre}
 L_a^{(\mu)}(y)=(-1)^a a!\int\limits_0^{2\pi}\frac{\dd\varphi}{2\pi} e^{\displaystyle-a i\varphi}(1+e^{i\varphi})^{\mu+a}e^{\displaystyle-ye^{i \varphi}}=y^a+\ldots
\end{equation}
Then the function appearing in the additional row and column for odd $k$, cf. Eq.~\eref{eq:partitionfunctionkodd}, is
\begin{eqnarray}
\mathcal F_{l}(\kappa_a,t)&=&(-1)^l\frac{l(l-1)Z_{l-2,\ga}(t)}{Z_{l,\ga}(t)}\biggl(L_{l-2}^{(2\gamma)}(\kappa_a)\label{func-2}\\
&&-(l-2)\frac{{\rm U}[\gamma+(l-1)/2,\gamma+1/2,t/2]}{{\rm U}[\gamma+(l-1)/2,\gamma+3/2,t/2]}L_{l-3}^{(2\gamma+1)}(\kappa_a)\biggl).\nonumber
\end{eqnarray}
The function was normalized via the known expansion 
to leading order in $\kappa_a$
\begin{equation}\mathcal F_{l}(\kappa_a,t)=(-1)^l l(l-1)Z_{l-2,\ga}(t)\kappa_a^{l-2}/Z_{l,\ga}(t)+\ldots\end{equation}
From the expression~\eref{func-2} we can readily read off the polynomials of even order ($l\to 2j+2$),
\begin{eqnarray}
\fl R_{2j}^{(\ga)}(y,t)&=& L_{2j}^{(2\gamma)}(y)-2j\frac{{\rm U}[j+\gamma+1/2,\gamma+1/2,t/2]}{{\rm U}[j+\gamma+1/2,\gamma+3/2,t/2]}L_{2j-1}^{(2\gamma+1)}(y)\nonumber\\
\fl&=& \frac{{\rm U}[j+\gamma+1/2,\gamma+1/2,t/2]}{{\rm U}[j+\gamma+1/2,\gamma+3/2,t/2]}L_{2j}^{(2\gamma+1)}(y)\nonumber\\
\fl&&+ \frac{2j+2\ga+1}{2}\frac{{\rm U}[j+\gamma+3/2,\gamma+3/2,t/2]}{{\rm U}[j+\gamma+1/2,\gamma+3/2,t/2]}L_{2j}^{(2\gamma)}(y),\label{pol-2}
\end{eqnarray}
for any integer $j\geq1$ and $R_{0}^{(\ga)}(y,t)=1$.
The second expression is the one presented in Ref.~\cite{AGKWWprl} and can be found by splitting the term $(1+r+e^{i\varphi})$ in $(1+e^{i\varphi})$ and $r$ instead off $e^{i\varphi}$ and $(1+r)$.

The odd polynomials are determined by their relation~\eref{eq:polynomialsodddegree} to the polynomials of even order. For this purpose we recall some recurrence relations of the monic Laguerre polynomials and the Tricomi confluent hypergeometric functions,
\begin{eqnarray}
  \fl\quad\quad\quad  &\left(y-2y\frac{\partial}{\partial y}\right)L_a^{(\mu)}(y)=L_{a+1}^{(\mu)}(y)+(\mu+1)L_a^{(\mu)}(y)-a(\mu+a)L_{a-1}^{(\mu)}(y),\\
  \fl\quad\quad\quad  &t\frac{\partial}{\partial t}{\rm U}\left(a,b,\frac{t}{2}\right)=(1-b){\rm U}\left(a,b,\frac{t}{2}\right)+(b-a-1){\rm U}\left(a,b-1,\frac{t}{2}\right).
\end{eqnarray}
 These relations yield
\begin{eqnarray}
\fl\quad\quad R_{2j+1}^{(\ga)}(y,t)&=&L_{2j+1}^{(2\gamma)}(y)+ (2\gamma+1+\widehat{c}_j(t))L_{2j}^{(2\gamma)}(y)-4j(\gamma+j)L_{2j-1}^{(2\gamma)}(y)\nonumber\\
\fl&&+ d_j^{(1)}(t)L_{2j}^{(2\gamma+1)}(y)+ d_j^{(2)}(t)L_{2j-1}^{(2\gamma+1)}(y)+ d_j^{(3)}(t)L_{2j-2}^{(2\gamma+1)}(y)\label{pol-3}
\end{eqnarray}
with the coefficients
\begin{eqnarray}
 \fl d_j^{(1)}(t)&=&-2j\frac{{\rm U}[j+\gamma+1/2,\gamma+1/2,t/2]}{{\rm U}[j+\gamma+1/2,\gamma+3/2,t/2]},\nonumber\\
  \fl d_j^{(2)}(t)&=&(2\gamma+\widehat{c}_j(t))d_j^{(1)}(t)+\left(d_j^{(1)}(t)\right)^2-4j(j+1)\frac{{\rm U}[j+\gamma+1/2,\gamma-1/2,t/2]}{{\rm U}[j+\gamma+1/2,\gamma+3/2,t/2]},\nonumber\\
  \fl d_j^{(3)}(t)&=&-2(2j-1)(\gamma+j)d_j^{(1)}(t).
\end{eqnarray}
We have already identified  part of the terms in $d_j^{(2)}(t)$ and $d_j^{(3)}(t)$ with the coefficient $d_j^{(1)}(t)$.

We underline that the ambiguous constant $c_j(t)$ for the odd polynomials, cf. Eq.~\eref{eq:polynomialsodddegree}, is not fixed, yet. Thus we are free to choose the coefficient $c_j(t)=-2\gamma-1$ such that one of the polynomials drops out in \eref{pol-3}.
Then the polynomials of odd order are a linear combination of only five Laguerre polynomials. This simplifies the result presented in Ref.~\cite{AGKWWprl}. Nevertheless we emphasize that both results are correct due to the various relations satisfied by the modified Laguerre polynomials and Tricomi's confluent hypergeometric functions, and the ambiguity in the constant $c_j(t)$ which was chosen differently in Ref.~\cite{AGKWWprl} compared to the simpler choice here. In the microscopic origin limit performed in section~\ref{sec:AsymptoticExpressions} we choose another constant $c_j(t)$ to simplify the asymptotic result.

The polynomials $\widehat{R}_j^{(\gamma)}(y,t)$, needed for the case $p=2L+1$ odd, can be obtained from the polynomials $R_j^{(\gamma)}(y,t)$ with the help of the relations~\eref{polynomial-rel}. Therefore the polynomials of even order, $\widehat{R}_{2j}^{(\gamma)}(y,t)$, are a linear combination of four Laguerre polynomials and the polynomials of odd order, $\widehat{R}_{2j+1}^{(\gamma)}(y,t)$, can be expressed as a sum of six Laguerre polynomials with a suitable choice of the constant $c_j(t)$.

%%%%%%%%%%%%%%%%%%%%%%%%%%%%%%%%%%%%%%%%%%%%%%%%%%%%%%%%%%%%%%%%%
\subsection{Kernel}\label{subsec:kernel}

The kernel $\mathcal K_{l}(\kappa_a,\kappa_b,t)$ can be first of all understood as a sum over the SOP, see Eq.~\eref{eq:Pfaffiankernel}. Plugging the results of subsection~\ref{subsec:polynomials} into this sum we are done.  However we can also start from the representation~\eref{kernel-av-1} and take the general result~\eref{calc-3} for $l\to l-2$, $k\to2$, and $\kappa\to \diag(\kappa_a,\kappa_b)\otimes\Id_2$. Then we find
\begin{eqnarray}\label{ker-1}
\fl\frac{\mathcal K_{l}(\kappa_a,\kappa_b,t)}{(\kappa_a-\kappa_b)}&\propto&\int\limits_{{\rm CSE}(4)} \dd\mu(U) {\det}^{-(l-2)/2}U{\det}^{(l+2\ga-2)/2}\left(\Id_{4}+U\right)\\
\fl&&\times \int_0^\infty\dd r \frac{r^{(l+2\ga-3)/2}}{(1+r)^{(l+2)/2}}{\det}^{1/2}\left([1+r]\Id_{4}+U\right)\exp\left(-\frac{1}{2}\left[\tr U\kappa+t\,r\right]\right).\nonumber
\end{eqnarray}
To evaluate the integrals we first expand the determinant coupling $U$ and $r$, i.e. 
\begin{equation}\fl\quad\quad\quad\quad{\det}^{1/2}\left([1+r]\Id_{4}+U\right)=(1+r)^2+(1+r)\tr U/2+{\det}^{1/2}U~,\end{equation}
and arrive at
\begin{eqnarray}\label{ker-2}
\fl&&\frac{\mathcal K_{l}(\kappa_a,\kappa_b,t)}{(\kappa_a-\kappa_b)}\\
\fl&\propto&\left[\int_0^\infty\dd r \frac{r^{(l+2\ga-3)/2}e^{-t\,r/2}}{(1+r)^{(l-2)/2}} -\int_0^\infty\dd r \frac{r^{(l+2\ga-3)/2}e^{-t\,r/2}}{(1+r)^{l/2}} \left(\frac{\partial}{\partial\kappa_a}+\frac{\partial}{\partial\kappa_b}\right)\right]\nonumber\\
\fl&&\times\int\limits_{{\rm CSE}(4)} \dd\mu(U) {\det}^{-(l-2)/2}U{\det}^{(l+2\ga-2)/2}\left(\Id_{4}+U\right)e^{-\tr U\kappa/2}\nonumber\\
\fl&&\hspace*{-0.5cm}+\int_0^\infty\dd r \frac{r^{(l+2\ga-3)/2}e^{-t\,r/2}}{(1+r)^{(l+2)/2}} \int\limits_{{\rm CSE}(4)} \dd\mu(U) {\det}^{-(l-3)/2}U{\det}^{(l+2\ga-2)/2}\left(\Id_{4}+U\right)e^{-\tr U\kappa/2}.\nonumber
\end{eqnarray}
The derivatives in $\kappa_a$ and $\kappa_b$ generate the trace of $U$.

Next we diagonalize the matrix $U=V\diag(e^{i\varphi_1},e^{i\varphi_2})\otimes \Id_2 V^\dagger=V\Phi V^\dagger$ with $V\in{\rm USp}(4)/{\rm USp}^2(2)$. The normalized measure becomes 
\begin{eqnarray}
\dd\mu(U)= |e^{i\varphi_1}-e^{i\varphi_2}|^4 \dd\varphi_1\dd\varphi_2\dd\mu_{\rm Haar}(V)/(6(2\pi)^2)
\end{eqnarray}
with $\dd\mu_{\rm Haar}(V)$ the normalized Haar measure on ${\rm USp}(4)/{\rm USp}^2(2)$. The integral over $V$ is an Itzykson-Zuber integral which is well-known~\cite{KoGu},
\begin{eqnarray}\label{IZ-bet4}
 \fl&&\int_{{\rm USp}(4)/{\rm USp}^2(2)}\dd\mu_{\rm Haar}(V)\exp\left[-\frac{1}{2}\tr V\Phi V^\dagger\kappa\right]\\
 \fl&=&6\left[\frac{(\kappa_a-\kappa_b)(e^{i\varphi_1}-e^{i\varphi_2})+2}{(\kappa_a-\kappa_b)^3(e^{i\varphi_1}-e^{i\varphi_2})^3}e^{-\kappa_a e^{i\varphi_1}-\kappa_b e^{i\varphi_2}}+\{\varphi_1\leftrightarrow\varphi_2\}\right].\nonumber
\end{eqnarray}
We plug this result into the integral over $U$ and have for two arbitrary $p,q\in\mathbb{N}$
\begin{eqnarray}
\fl&&\int\limits_{{\rm CSE}(4)} \dd\mu(U) {\det}^{-p/2}U{\det}^{(p+q)/2}\left(\Id_{4}+U\right)e^{-\tr U\kappa/2}\nonumber\\
\fl&=& \frac{2}{(\kappa_a-\kappa_b)^3}\int_0^{2\pi}\frac{d\varphi_1}{2\pi}\int_0^{2\pi}\frac{d\varphi_2}{2\pi}e^{-i(p+2)(\varphi_1+\varphi_2)}(1+e^{i\varphi_1})^{p+q}(1+e^{i\varphi_2})^{p+q}\nonumber\\
\fl&&\times\left[(\kappa_a-\kappa_b)(e^{i\varphi_1}-e^{i\varphi_2})^2+2(e^{i\varphi_1}-e^{i\varphi_2})\right]e^{-\kappa_a e^{i\varphi_1}-\kappa_b e^{i\varphi_2}}\nonumber\\
\fl&=&\frac{2}{(p+1)!(p+2)!}\frac{1}{(\kappa_a-\kappa_b)^3}\nonumber\\
\fl&&\times\left[(\kappa_a-\kappa_b)\left((p+1)L_p^{(q)}(\kappa_a)L_{p+2}^{(q-2)}(\kappa_b)-2(p+2)L_{p+1}^{(q-1)}(\kappa_a)L_{p+1}^{(q-1)}(\kappa_b)\right.\right.\nonumber\\
\fl&&\left.\left.+(p+1)L_{p+2}^{(q-2)}(\kappa_a)L_{p}^{(q)}(\kappa_b)\right)-2L_{p+1}^{(q-1)}(\kappa_a)L_{p+2}^{(q-2)}(\kappa_b)+2L_{p+2}^{(q-2)}(\kappa_a)L_{p+1}^{(q-1)}(\kappa_b)\right]\nonumber\\
\fl&=&\frac{2}{([p+2]!)^2}\left[\frac{1}{\kappa_a-\kappa_b}\left(\frac{\partial}{\partial\kappa_a}-\frac{\partial}{\partial\kappa_b}\right)\right]^2L_{p+2}^{(q-2)}(\kappa_a)L_{p+2}^{(q-2)}(\kappa_b).\label{coset-int-1}
\end{eqnarray}
The second equality results from an expansion of the polynomials in the bracket in the phases $e^{i\varphi_1}$ and $e^{i\varphi_2}$.  The third equality is a compact representation by rewriting the polynomial in the brackets as derivatives in $\kappa_a$ and $\kappa_b$. We remark that Eq.~\eref{coset-int-1} is normalized to the leading order term $2(\kappa_a\kappa_b)^p/[p!(p+2)!]$.

 We combine the intermediate result~\eref{coset-int-1}, the definition of Tricomi's confluent hypergeometric function~\eref{confluent}, and the derivative of monic Laguerre polynomials $\partial_y L_a^{(\mu)}(y)=a L_{a-1}^{(\mu+1)}(y)$. Then we find for the kernel
\begin{eqnarray}\label{ker-3}
\fl\mathcal K_{l}(\kappa_a,\kappa_b,t)&=&\frac{Z_{l-2,\ga}(t)}{Z_{l,\ga}(t)}(\kappa_a-\kappa_b)\left[\frac{1}{\kappa_a-\kappa_b}\left(\frac{\partial}{\partial\kappa_a}-\frac{\partial}{\partial\kappa_b}\right)\right]^2\\
\fl&&\times\left[1-\frac{\U\left(\gamma+(l-1)/2,\gamma+1/2,t/2\right)}{\U\left(\gamma+(l-1)/2,\gamma+3/2,t/2\right)}\left(\frac{\partial}{\partial\kappa_a}+\frac{\partial}{\partial\kappa_b}\right)\right.\nonumber\\
\fl&&\left.+\frac{\U\left(\gamma+(l-1)/2,\gamma-1/2,t/2\right)}{\U\left(\gamma+(l-1)/2,\gamma+3/2,t/2\right)}\frac{\partial^2}{\partial\kappa_a\partial\kappa_b}\right]L_{l}^{(2\gamma-2)}(\kappa_a)L_{l}^{(2\gamma-2)}(\kappa_b).\nonumber
\end{eqnarray}
 Although we have not identified the recurrence relation of the skew-orthogonal polynomials we have been able to derive the Christoffel-Darboux formula corresponding to the sum~\eref{eq:Pfaffiankernel}. It expresses the whole sum as a finite small number of terms. Each term is a product of two Laguerre polynomials. The total number of these terms is twelve after differentiating and ordering the Laguerre polynomials with respect to there index. This is a fixed number of terms which shows that this expression of the kernel is ideal to study the large $l$ behaviour including the asymptotic behaviour in the bulk and the soft edge scaling limit. Nonetheless we employ the expression as a sum over the skew-orthogonal polynomials, see Eq.~\eref{eq:Pfaffiankernel}, in the derivation of the gap probability and the distribution of the smallest eigenvalue in the microscopic limit. The reasons are the additional derivatives we have to perform resulting from degeneracy of the variables $\kappa_j\to-t$, see subsection~\ref{subsec:distribution}. Then the result~\eref{ker-3} becomes quite nasty due to a $1/(\kappa_a-\kappa_b)$ term in the differential operator in front of the product of the two Laguerre polynomials.

%%%%%%%%%%%%%%%%%%%%%%%%%%%%%%%%%%%%%%%%%%%%%%%%%%%%%%%%%%%%%%%%%

\subsection{Gap Probability and Distribution of the Smallest Eigenvalue}\label{subsec:distribution}

To obtain the gap probability or the distribution of the smallest eigenvalue itself we have to set all variables equal, $\kappa_1=\ldots=\kappa_k=-t$. Hence we have to apply l'H\^{o}pital's rule in Eqs.~\eref{eq:partitionfunctionkeven} and \eref{eq:partitionfunctionkodd} yielding
\begin{eqnarray}
\fl\left\langle \Det^k(X+t\Id_p)\right\rangle_{p,\ga}^t&=&
\frac{(-1)^{k(k-1)/2}p!Z_{p+k,\ga}(t)}{(p+k)!Z_{p,\ga}(t)\prod_{j=0}^{k-1}j!}
\label{eq:partitionfunctionkeven.b}\\
  \fl&&\times\Pf_{1\leq a,b \leq k}\left[\partial_{\kappa_1}^{a-1}\partial_{\kappa_2}^{b-1}\mathcal K_{p+k}(\kappa_1,\kappa_2,t)|_{\kappa_1=\kappa_2=-t}  \right]\nonumber
\end{eqnarray}
for $k=2m$ even and
\begin{eqnarray}
\fl\left\langle \Det^k(X+t\Id_p)\right\rangle_{p,\ga}^t&=&\frac{(-1)^{k(k-1)/2}p!Z_{p+k+1,\ga}(t)}{(p+k+1)!Z_{p,\ga}(t)\prod_{j=0}^{k-1}j!}
 \label{eq:partitionfunctionkodd.b}\\
\fl&&\hspace*{-3cm}\times
\Pf_{1\leq a,b \leq k}\left[
\begin{array}{ll}
\partial_{\kappa_1}^{a-1}\partial_{\kappa_2}^{b-1}\mathcal K_{p+k+1}(\kappa_1,\kappa_2,t)|_{\kappa_1=\kappa_2=-t} & \partial_{\kappa}^{a-1}\mathcal F_{p+k+1}(\kappa,t)|_{\kappa=-t}\\
-\partial_{\kappa}^{b-1}\mathcal F_{p+k+1}(\kappa,t)|_{\kappa=-t} &0\\
\end{array}\nonumber
\right]
\end{eqnarray}
for $k=2m+1$ odd. The additional sign and the product of inverse factorials in front of the Pfaffians result from differentiating the Vandermonde determinant $\Delta_k(\kappa)$, cf. Eqs.~\eref{eq:partitionfunctionkeven} and \eref{eq:partitionfunctionkodd}.

The prefactor in front of the averages for the kernels $K_{p+k}(\kappa_1,\kappa_2,t)$ and polynomials $F_{p+k+1}(\kappa,t)$, 
have to be considered together with the prefactors in 
Eqs.~\eref{kernel-av-1} and \eref{kernel-av-2}.
Hence we can normalize the two kernels such that the microscopic origin limit of these kernels is finite in preparation of section~\ref{sec:AsymptoticExpressions}. We find the result for arbitrary $\gamma$,
\begin{eqnarray}
\fl\left\langle \Det^k(X+t\Id_p)\right\rangle_{p,\ga}^t&=&
C_{pk}^{(\gamma)}(t)\Pf_{0\leq a,b \leq k-1}\left[\Xi_{ab}^{(\gamma,p+k)}(t)  \right]
\label{eq:partitionfunctionkeven.c}
\end{eqnarray}
for $k=2m$ even and
\begin{eqnarray}
\fl\left\langle \Det^k(X+t\Id_p)\right\rangle_{p,\ga}^t&=&C_{pk}^{(\gamma)}(t)\Pf_{0\leq a,b \leq k-1}\left[
\begin{array}{ll}
\Xi_{ab}^{(\gamma,p+k+1)}(t) & \xi_{a}^{(\gamma,p+k+1)}(t)\\
-\xi_{b}^{(\gamma,p+k+1)}(t) &0\\
\end{array}\right]\label{eq:partitionfunctionkodd.c}
\end{eqnarray}
for $k=2m+1$ odd. The functions serving as the kernels are
\begin{eqnarray}\label{kernel-K}
\fl \Xi_{ab}^{(\gamma,l)}(t)&=&\frac{(-1)^{a+b}t^{2\gamma+a+b+1}l(l-1)Z_{l-2,\gamma}(t)}{Z_{l,\gamma}(t)}\partial_{\kappa_1}^{a}\partial_{\kappa_2}^{b}(\kappa_1-\kappa_2)\\
\fl&&\times\left\langle \Det\left(X-\kappa_1\Id_{l-2}\right)\Det\left(X-\kappa_2\Id_{l-2}\right)\right\rangle_{l-2,\ga}^{t}|_{\kappa_1=\kappa_2=-t}\nonumber\\
\fl&=&(-1)^{a+b}t^{2\gamma+a+b+1}\nonumber\\
\fl&&\hspace*{-1.5cm}\times\left\{\begin{array}{cl} \displaystyle\sum_{j=0}^{(l-2)/2} \frac{\partial_{\kappa_1}^{a}R_{2j+1}^{(\ga)}\left(\kappa_1,t\right)\partial_{\kappa_2}^{b}R_{2j}^{(\ga)}\left(\kappa_2,t\right)
-\partial_{\kappa_2}^{b}R_{2j+1}^{(\ga)}\left(\kappa_2,t\right)
\partial_{\kappa_1}^{a}R_{2j}^{(\ga)}\left(\kappa_1,t\right)}{r_j^{(\ga)}(t)}, & \\ &\hspace*{-1cm} l\in2\mathbb{N}, \\
\displaystyle \sum_{j=0}^{(l-3)/2} \frac{\partial_{\kappa_1}^{a}\widehat{R}_{2j+1}^{(\ga)}\left(\kappa_1,t\right)\partial_{\kappa_2}^{b} \widehat{R}_{2j}^{(\ga)}\left(\kappa_2,t\right)
-\partial_{\kappa_2}^{b}\widehat{R}_{2j+1}^{(\ga)}\left(\kappa_2,t\right)
\partial_{\kappa_1}^{a}\widehat{R}_{2j}^{(\ga)}\left(\kappa_1,t\right)}{r_j^{(\ga)}(t)}, & \\ &\hspace*{-1cm}  l\in2\mathbb{N}+1, \end{array}\right.\nonumber
\end{eqnarray}
and
\begin{eqnarray}\label{kernel-F}
\fl \xi_{a}^{(\gamma,l)}(t)&=&\frac{(-1)^{a}t^{2\gamma+a}}{(l-2)!}\partial_{\kappa}^{a}\left\langle \Det\left(X-\kappa\Id_{l-2}\right)\right\rangle_{l-2,\ga}^{t}|_{\kappa=-t}\\
\fl&=&(-1)^{a+l}t^{2\gamma+a}\biggl(\frac{L_{l-a-2}^{(2\gamma+a)}(-t)}{(l-a-2)!}-\frac{{\rm U}[\gamma+(l-1)/2,\gamma+1/2,t/2]}{{\rm U}[\gamma+(l-1)/2,\gamma+3/2,t/2]}\frac{L_{l-a-3}^{(2\gamma+a+1)}(-t)}{(l-a-3)!}\biggl).\nonumber
\end{eqnarray}
Due to this particular normalization we have the overall constant
\begin{eqnarray}\label{kernel-finite-N}
C_{pk}^{(\gamma)}(t)&=&\left\{\begin{array}{cl}
 \displaystyle\frac{p!}{(p+k)!\prod_{j=0}^{k-1}j!}\frac{Z_{p+k,\ga}(t)}{t^{k(k+2\gamma)/2}Z_{p,\ga}(t)}, & k\in 2\mathbb{N}_0,\\
  \displaystyle \frac{p!}{\prod_{j=0}^{k-1}j!}\frac{Z_{p+k-1,\ga}(t)}{t^{k(k+2\gamma)/2+(2\gamma-1)/2}Z_{p,\ga}(t)},  & k\in2\mathbb{N}_0+1.
\end{array}\right.
\end{eqnarray}
Note that we have not included the prefactors shown in Eqs.~\eref{gapvevt} and \eref{1stvev.b} for the full expressions of the gap probability and the distribution of the smallest eigenvalue. 
For the definition of the limiting quantities see Eqs. \eref{e-def} and \eref{p-def} where all factors in $p$ are accounted for.

First we show the explicit expression of the gap probability at finite $N$. We multiply the results~\eref{eq:partitionfunctionkeven.c} and \eref{eq:partitionfunctionkodd.c} for $\gamma=0$ with the factor $e^{-pt/2}Z_{p,0}(t)/Z_{p,2k}$. Then we find the first of our main results,
\begin{eqnarray}
  \fl E_{p,2k}(t) &=&\mathcal C_{p,k}^{(0)}
 (4p t)^{-k^2/2+1/2}e^{-pt/2} \frac{\Gamma[(p+k+1)/2]\text{U}\left((p+k+1)/2,3/2;t/2\right)}{2\sqrt{2p}}\nonumber\\
 \fl&&\times\Pf_{0\leq a,b \leq k-1}\left[  \Xi_{ab}^{(0,p+k)}(t)\right]
  \label{eq:GapPfaffEvenAlpha}
\end{eqnarray}
for even $k$ and
\begin{eqnarray}
\fl E_{p,2k}(t) &=&  \mathcal C_{p,k}^{(0)}(4p t)^{-k^2/2+1}e^{-pt/2} \frac{\Gamma[(p+k)/2]\text{U}\left((p+k)/2,3/2;t/2\right)}{2\sqrt{2p}}\nn\\
\fl&&\times\Pf_{0\leq a,b \leq k-1}\left[
\begin{array}{ll}
\Xi_{ab}^{(0,p+k+1)}(t) & \xi_{a}^{(0,p+k+1)}(t)\\
-\xi_{b}^{(0,p+k+1)}(t) &0\\
\end{array}\right]
\label{eq:GapPfaffOddAlpha}
\end{eqnarray}
for odd $k$, where we introduce the global normalization constant
% The constant,
% \begin{eqnarray}
% \label{eq:GlobalNormalizationConstant.a}
% \fl\mathcal C_{p,k}^{(0)}= \left\{\begin{array}{cl} \frac{2^{3k^2/2}p!p^{k^2/2}}{(p+k)!\Gamma[(p+k+1)/2]\prod_{j=0}^{k-1}j!}\frac{\prod_{j=0}^{p+k-1}\Gamma[(j+3)/2]\Gamma[(j+2)/2]/\Gamma[3/2]}{\prod_{j=0}^{p-1}\Gamma[(j+3)/2]\Gamma[(j+2k+1)/2]/\Gamma[3/2]}, & k\in2\mathbb{N},\\ 
% \frac{2^{(3k^2-2k-2p-1)/2}p!p^{(k^2-1)/2}}{\Gamma[(p+k)/2]\prod_{j=0}^{k-1}j!}\frac{\prod_{j=0}^{p+k-2}\Gamma[(j+3)/2]\Gamma[(j+2)/2]/\Gamma[3/2]}{\prod_{j=0}^{p-1}\Gamma[(j+3)/2]\Gamma[(j+2k+1)/2]/\Gamma[3/2]}, & k\in2\mathbb{N}+1,  \end{array}\right.
% \end{eqnarray}
%  is relatively complicated but it is chosen $t$ independent and remains finite in the microscopic limit. The constant can be simplified by employing the duplication formula of the Gamma function and the following two identities
% \begin{eqnarray}
%  \Gamma\left[\frac{j+k+3}{2}\right]\Gamma\left[\frac{j+k+2}{2}\right]&=&\Gamma\left[\frac{j+3}{2}\right]\Gamma\left[\frac{j+2}{2}\right]\prod_{l=1}^k\frac{j+l+1}{2},\nonumber\\
%  \frac{1}{\Gamma[(p+1)/2]}\prod\limits_{j=1}^{p-1}\frac{\Gamma[(j+2)/2]}{\Gamma[(j+1)/2]}&=&\frac{1}{\sqrt{\pi}}.\label{helpful}
% \end{eqnarray}
% Then the constant becomes
\begin{equation}
\label{eq:GlobalNormalizationConstant.b}
\eqalign{\mathcal C_{p,k}^{(0)}&=\prod\limits_{l=0}^{k-1}\frac{4^{l+1}(2l)!\Gamma[p+l+2]p^{l-1}}{l!\Gamma[p+2l+1]}  
\\&\times\left\{\begin{array}{cl} \displaystyle\frac{1}{\sqrt{\pi}}\frac{2^{-k/2}p!p^{3k/2}\Gamma[(p+1)/2]}{(p+k)!\Gamma[(p+k+1)/2]} &,
k\in2\mathbb{N},\\ 
\displaystyle\frac{1}{\sqrt{\pi}}\frac{2^{-(k+3)/2}p!p^{(3k-1)/2}\Gamma[(p+1)/2]}{(p+k)!\Gamma[(p+k)/2]}&,  k\in2\mathbb{N}+1  \end{array}\right.}
\end{equation}
which is also ideal to take the limit $p\to\infty$.

\begin{figure}
 \centerline{\includegraphics[width=0.49\textwidth]{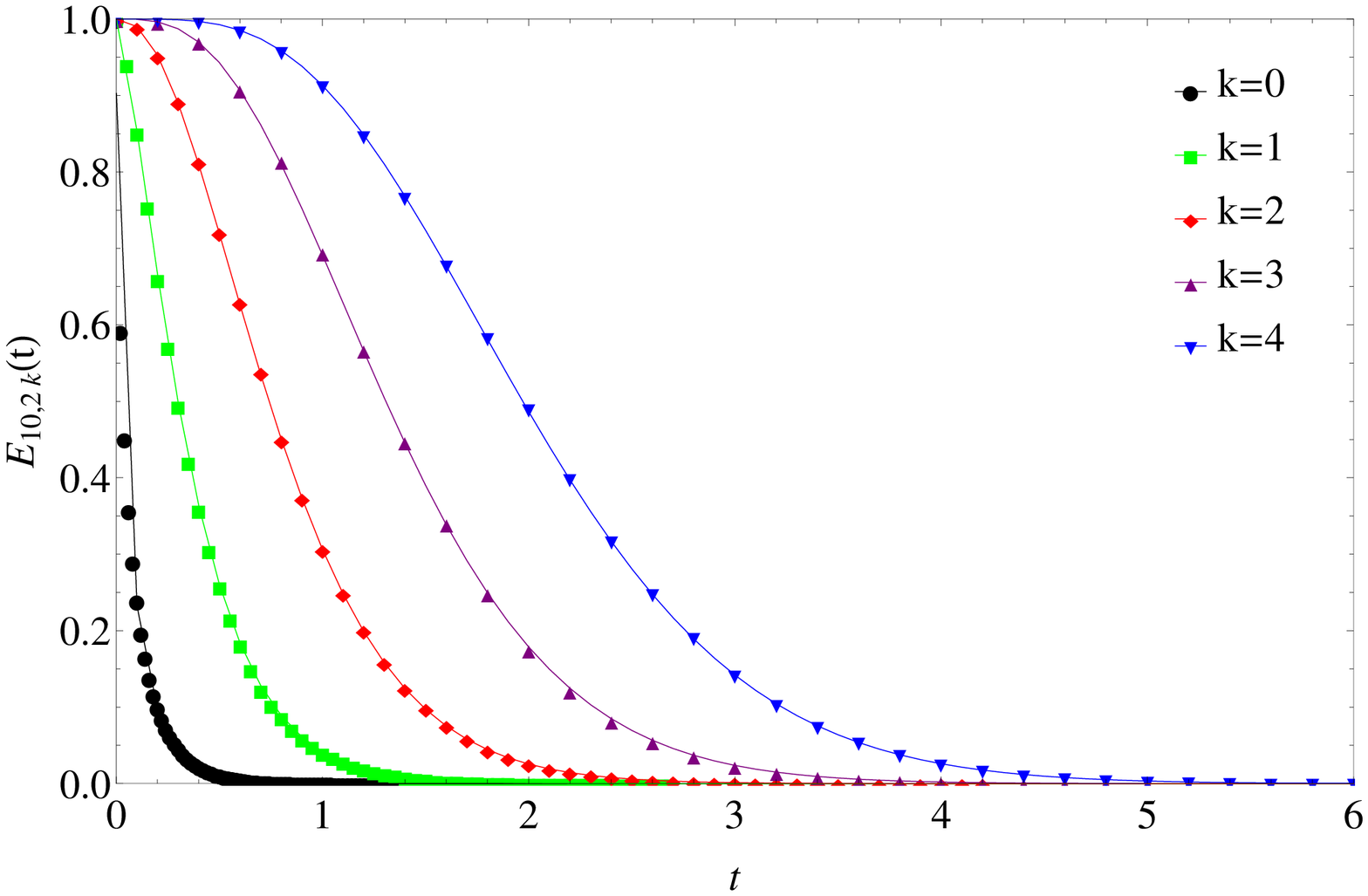}\hfill\includegraphics[width=0.49\textwidth]{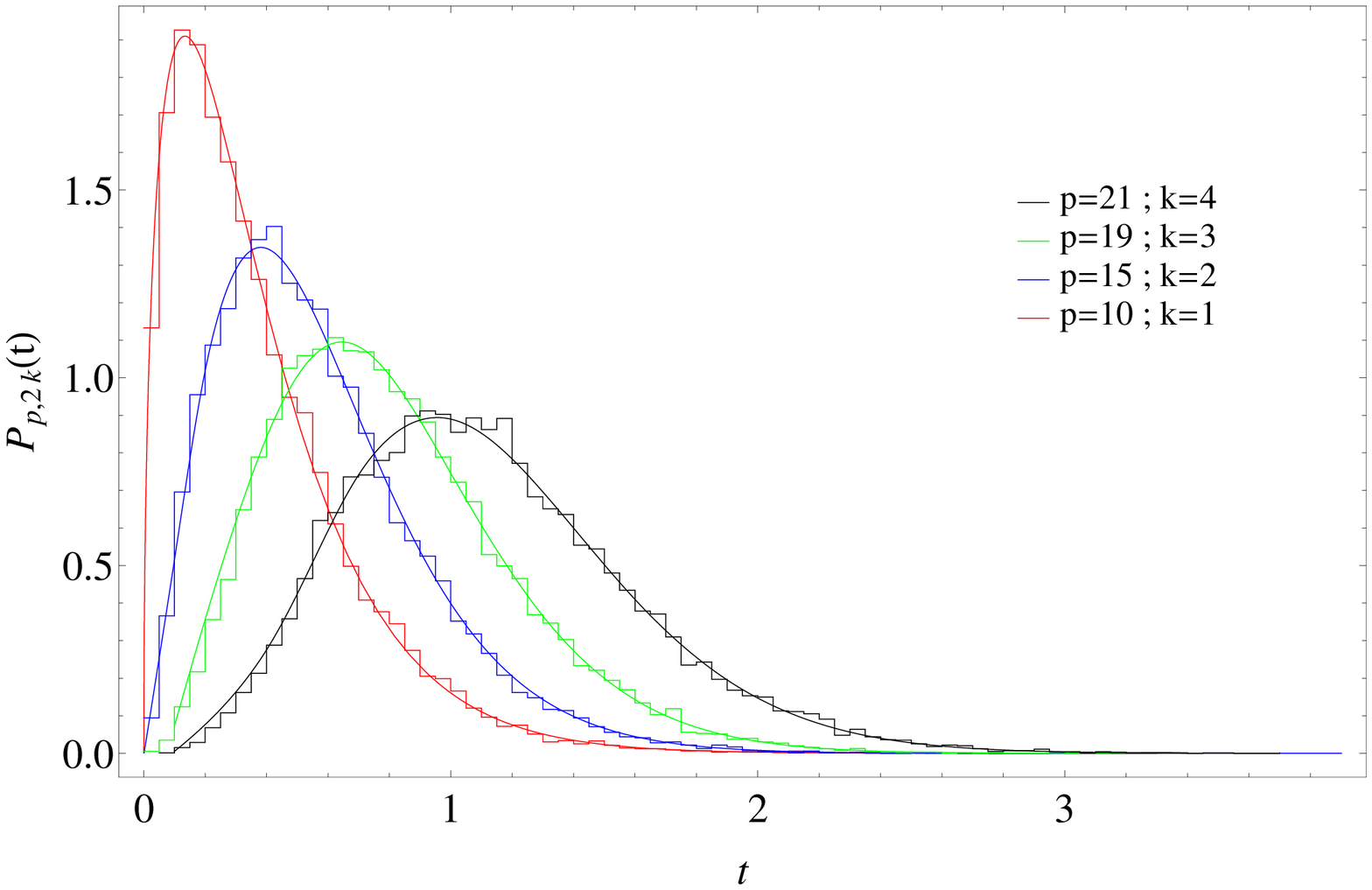}}
 % Egap.png: 1199x790 pixel, 72dpi, 42.29x27.87 cm, bb=0 0 1199 790
 \caption{Visualization of our analytical results~\eref{eq:GapPfaffEvenAlpha}, \eref{eq:GapPfaffOddAlpha}, \eref{eq:smallDist:FinalOddpEvenAlpha}, and \eref{eq:smallDist:FinalOddpOddAlpha} for finite $p=10$ (solid curves) compared to Monte Carlo simulations (symbols, histogram) for the gap probability (left plot) and the distribution of the smallest eigenvalue (right plot). For the gap probability we generated $10000$ real Wishart matrices of size $10\times (10+\nu)$ with an index $\nu=2k=0,2,4,6,8$.
For the distribution of the smallest eigenvalue
we simulated $20000$ random matrices with various dimensions $p$ shown in the inset corresponding to $\nu=2k=2,4,6,8$.
}
 \label{fig:finite-p}
 \end{figure}

The explicit expression of the kernels $\Xi_{ab}^{(0,p+k)}(t)$ and $\xi_{b}^{(0,p+k+1)}(t)$ are not much more enlightening at finite $p$ than the expressions~\eref{kernel-K} and \eref{kernel-F} for general $\gamma$. Therefore we skip their expressions here and show them explicitly for the particular case $\gamma=0$ and $\gamma=1$ in the microscopic limit in section~\ref{sec:AsymptoticExpressions}. For finite $p$ we visualize the gap probability in Fig.~\ref{fig:finite-p}.

We skip the explicit kernels for the distribution of the smallest eigenvalue, too, and only explicitly show the Pfaffian structure with its normalization constant. For this reason we multiply the results~\eref{eq:partitionfunctionkeven.c} and \eref{eq:partitionfunctionkodd.c} for $\gamma=1$ and $p\to p-1$ with the factor $pt^{(2k-1)/2}e^{-pt/2}Z_{p-1,1}(t)/Z_{p,2k}$ from Eq. \eref{1stvev.b}
\begin{eqnarray}
  \fl P_{p,2k}(t) \dd t&=&\mathcal C_{p,k}^{(1)}
  (4pt)^{-k^2/2+1}e^{-pt/2} \frac{\Gamma[(p+k+2)/2]\text{U}\left((p+k+2)/2,5/2;t/2\right)}{2(2p)^{3/2}}\nonumber\\
  \fl&&\times\Pf_{0\leq a,b \leq k-1}\left[  \Xi_{ab}^{(1,p+k-1)}(t)\right]\dd(4pt)
\label{eq:smallDist:FinalOddpEvenAlpha}
\end{eqnarray}
for even $k$ and
\begin{eqnarray}
\fl P_{p,2k}(t) \dd t&=&  \mathcal C_{p,k}^{(1)}(4pt)^{-k^2/2+1/2}e^{-pt/2} \frac{\Gamma[(p+k+1)/2]\text{U}\left((p+k+1)/2,5/2;t/2\right)}{2(2p)^{3/2}} \nn\\
\fl&&\times\Pf_{0\leq a,b \leq k-1}\left[
\begin{array}{ll}
\Xi_{ab}^{(1,p+k)}(t) & \xi_{a}^{(1,p+k)}(t)\\
-\xi_{b}^{(1,p+k)}(t) &0\\
\end{array}\right]\dd(4pt)
\label{eq:smallDist:FinalOddpOddAlpha}
\end{eqnarray}
for odd $k$. We multiplied the differential $\dd t$ to emphasize that this quantity is a density and transforms as a pseudo scalar under changes of coordinates. The constant 
\begin{eqnarray}
\label{eq:smallDist:OverallNormalization}
\eqalign{\mathcal C_{p,k}^{(1)}%&=& \left\{\begin{array}{cl} \frac{2^{(3k^2-7)/2}p!p^{(k^2-1)/2}}{(p+k-1)!\Gamma[(p+k+2)/2]\prod_{j=0}^{k-1}j!}\frac{\prod_{j=0}^{p+k-2}\Gamma[(j+3)/2]\Gamma[(j+4)/2]/\Gamma[3/2]}{\prod_{j=0}^{p-1}\Gamma[(j+3)/2]\Gamma[(j+2k+1)/2]/\Gamma[3/2]}, \ 
%& \\ & \hspace*{-0.5cm}  
% k\in2\mathbb{N},\\ 
% & \\ 
% \frac{2^{(3k^2-2k-2p-4)/2}p!p^{k^2/2}}{\Gamma[(p+k+1)/2]\prod_{j=0}^{k-1}j!}\frac{\prod_{j=0}^{p+k-3}\Gamma[(j+3)/2]\Gamma[(j+4)/2]/\Gamma[3/2]}{\prod_{j=0}^{p-1}\Gamma[(j+3)/2]\Gamma[(j+2k+1)/2]/\Gamma[3/2]},\  
%& \\ & \hspace*{-0.5cm}  
% k\in2\mathbb{N}+1  \end{array}\right.\nonumber\\
% \fl
&=\prod\limits_{l=0}^{k-1}\frac{4^{l+1}(2l)!\Gamma[p+l+2]p^{l-1}}{l!\Gamma[p+2l+1]}
\\&\times \left\{\begin{array}{cl} \displaystyle\frac{1}{\sqrt{\pi}}\frac{2^{-(k+5)/2}p!p^{(3k-1)/2}\Gamma[(p+1)/2]}{(p+k)!\Gamma[(p+k)/2]}&,
k\in2\mathbb{N}\\ 
\displaystyle\frac{1}{\sqrt{\pi}}\frac{2^{-(k+6)/2}p!p^{3k/2}\Gamma[(p+1)/2]}{(p+k)!\Gamma[(p+k+1)/2]}&,  k\in2\mathbb{N}+1  \end{array}\right.}
\end{eqnarray}
% This constant was calculated in a similar way as the one in Eq.~\eref{eq:GlobalNormalizationConstant.b}. Also this constant
it is $t$ independent and converges to a finite number in the limit $p\to\infty$.

Equations~\eref{eq:smallDist:FinalOddpEvenAlpha} and \eref{eq:smallDist:FinalOddpOddAlpha} are our second main result. The distributions at finite $p$ are visualized in Fig.~\ref{fig:finite-p}. The two simplest cases for $k=0$,
\begin{equation}\label{k=0-p-finite}
P_{p,0}(t)=\frac{p!}{2^{p-1/2}\Gamma[p/2]}\frac{1}{\sqrt{t}}e^{-pt/2}\U\left(\frac{p-1}{2},-\frac{1}{2},\frac{t}{2}\right),
\end{equation}
 and $k=1$
\begin{eqnarray}\label{k=1-p-finite}
\fl P_{p,2}(t)&=&\frac{\Gamma[(p+1)/2]}{\sqrt{2\pi}}\sqrt{t}e^{-pt/2}\\
\fl&&\hspace*{-1cm}\times\left[\U\left(\frac{p-1}{2},-\frac{1}{2},\frac{t}{2}\right)\frac{(-1)^{p-1}L_{p-1}^{(2)}(-t)}{(p-1)!}+\U\left(\frac{p+1}{2},\frac{1}{2},\frac{t}{2}\right)\frac{(-1)^{p-2}L_{p-2}^{(3)}(-t)}{(p-2)!}\right],\nonumber
\end{eqnarray}
obviously agree with the results by Edelman~\cite{Edelman}.  We underline that he has employed the standard normalization of the Laguerre polynomials while we have chosen the monic normalization. Moreover we have used Kummer's identity of Tricomi's confluent hypergeometric function and the duplication formula of the Gamma function to obtain Eqs.~\eref{k=0-p-finite} and \eref{k=1-p-finite} from Eqs.~\eref{eq:smallDist:FinalOddpEvenAlpha} and \eref{eq:smallDist:FinalOddpOddAlpha}, respectively.

%%%%%%%%%%%%%%%%%%%%%%%%%%%%%%%%%%%%%%%%%%%%%%%%%%%%%%%%%%%%%%%%%%%%

\section{Microscopic Origin Limit}\label{sec:AsymptoticExpressions}

In the microscopic origin limit of the gap probability~\eref{eq:definitionE} we have to perform the limit  $p\rightarrow\infty$ while keeping $2k=\nu=n-p$ and $t=u/4p$ fixed. Hence we zoom into a region of  scale $1/p$ around the origin. In this region chiral random matrix theory is identical with physical theories like QCD below the critical temperature \cite{Jac,JacTilo} and condensed matter theory of disordered system for particular topological insulators and superconductors~\cite{carlo}. In this regime the limiting gap probability and distribution of the first eigenvalue are defined as follows:
\begin{equation}\label{e-def}
\mathcal{E}_{\nu}(u) := \lim_{p\rightarrow\infty}E_{p,\nu}\left(\frac{u}{4p}\right)
\end{equation}
and
\begin{equation}\label{p-def}
\mathcal{P}_{\nu}(u) \dd u:= \lim_{p\rightarrow\infty}P_{p,\nu}\left(\frac{u}{4p}\right)\dd\left(\frac{u}{4p}\right).
\end{equation}
We underline that the scaling factor resulting from the differential in the 
second definition is crucial to obtain a finite limit.

The first question we have to address is the dependence of the limit on the parity of $p$ (if $p$ is even or odd) because the skew-orthogonal polynomials crucially depend on it, cf. Eqs.~\eref{eq:definitionorhtogonalpoly} and \eref{eq:additionalrequiermentpodd}. We can circumvent this problem by the fact that all important quantities can be written  in terms of averages of characteristic polynomials to some powers, in particular the normalization~\eref{part-norm} and the kernels~\eref{kernel-av-1} and \eref{kernel-av-2}. Therefore it is sufficient to show that the limit of the average~\eref{average} is independent of the parity of $p$. For this reason we  consider the intermediate result~\eref{calc-3}. The microscopic limit can be readily performed by rescaling the integral $U\to 2p U$ and $r\to 2p r$ yielding 
\begin{eqnarray}
\fl \lim_{p\to\infty} I_p\left(\frac{\kappa}{4p}\right)&\propto& \int\limits_{{\rm CSE}(2k)} \dd\mu(U) {\det}^{\ga}U\int_0^\infty\dd r r^{(2\ga-2k-1)/2}{\det}^{1/2}\left(r\Id_{2k}+U\right)\nonumber\\
\fl&&\times\exp\left(-\frac{1}{4}\left[\tr U\kappa-\tr U^{-1}+t\,r+r^{-1}\right]\right),\label{calc-asymp}
\end{eqnarray}
up to a $p$ independent normalization constant. In particular this result is independent if the limit has been approached by an even $p$ or an odd one. Therefore we do the asymptotic analysis for even $p$, only. 

The behaviour of the Laguerre polynomials and Tricomi's confluent hypergeometric function in the microscopic limit determines the whole asymptotics. The asymptotics of both kinds of functions are given by
\begin{equation}
 \label{eq:MicroscopicLimitL}
\lim\limits_{p\to\infty}\frac{(-1)^{ap+c} p^{-b} }{\Gamma\left(ap +c+1\right)}L_{ap +c}^{(b)}\left(-\frac{u}{4p}\right) = \left(\frac{4a}{u}\right)^{b/2} I_{b}\left(\sqrt{au}\right),
\end{equation}
and
\begin{equation}
\label{eq:MicroscopicLimitU}
 \hspace*{-1cm}\lim\limits_{p\to\infty}\frac{\Gamma\left(ap +c\right)}{p^{b-1}}\text{U}\left(ap+c,b;\frac{u}{8p}\right)= 2\left(\frac{8a }{u}\right)^{(b-1)/2} K_{b-1}\left(\sqrt{\frac{au}{2}}\right),
\end{equation}
respectively. Here we employ the modified Bessel functions of the first and second kind, $I_a$ and $K_a$, respectively. The auxiliary parameters $a,b,c$ take certain values in the particular limits below.

The limits~\eref{eq:MicroscopicLimitL} and \eref{eq:MicroscopicLimitU} imply the following asymptotics of the derivatives of the polynomials for even order
\begin{eqnarray}
 \fl\xi_a^{(\gamma,\infty)}\left(xu\right)&:=&\lim\limits_{p\to\infty} \xi_a^{(\gamma,xp)}\left(\frac{u}{4p}\right)
%\nonumber\\
% \fl&=&
=\left.\lim\limits_{p\to\infty}\frac{(-1)^{xp-a}}{\Gamma[xp+1]}\left(\frac{u}{4p}\right)^{2\gamma+a}\partial_\kappa^{a}R_{xp}^{(\gamma)}\left(\kappa,\frac{u}{4p}\right)\right|_{\kappa=-u/(4p)}\nonumber\\ \fl&=&\left(\frac{xu}{4}\right)^{(2\gamma+a)/2}\left[I_{2\gamma+a}(\sqrt{xu})+\frac{K_{\gamma-1/2}(\sqrt{xu/4})}{K_{\gamma+1/2}(\sqrt{xu/4})}I_{2\gamma+a+1}(\sqrt{xu})\right]\label{lim-pol-even}
\end{eqnarray}
and for odd order
\begin{eqnarray}
 \fl&&\left.\lim\limits_{p\to\infty}\frac{(-1)^{xp-a}}{\Gamma[xp+1]}\left(\frac{u}{4p}\right)^{2\gamma+a}\partial_\kappa^{a}R_{xp+1}^{(\gamma)}\left(\kappa,\frac{u}{4p}\right)\right|_{\kappa=-u/(4p)}\nonumber\\
 \fl&=&\lim\limits_{p\to\infty}\frac{(-1)^{xp-a}}{\Gamma[xp+1]}\left(\frac{u}{4p}\right)^{2\gamma+a}\nonumber\\
 \fl&&\times\left.\partial_\kappa^{a}\left(\kappa+\widehat{c}_{px/2}\left(\frac{u}{4p}\right)-2 \kappa\partial_\kappa-2 u\partial_u \right)R_{xp}\left(\kappa,\frac{u}{4p}\right)\right|_{\kappa=-u/(4p)}\nonumber\\
 \fl&=&\left[C(xu)-2a-2u\partial_u\right]\lim\limits_{p\to\infty}\xi_a^{(\gamma,xp)}\left(\frac{u}{4p}\right)\nonumber\\
 \fl&=&\left(\frac{xu}{4}\right)^{(2\gamma+a)/2}\left[I_{2\gamma+a}(\sqrt{xu})\left(C(xu)-2a-\sqrt{xu}\frac{K_{\gamma-1/2}(\sqrt{xu/4})}{K_{\gamma+1/2}(\sqrt{xu/4})}\right)+I_{2\gamma+a+1}(\sqrt{xu})\right.\nonumber\\
 \fl&&\left.\times\left(\left[C(xu)+2\gamma+1-\sqrt{\frac{xu}{4}}\frac{K_{\gamma-1/2}(\sqrt{xu/4})}{K_{\gamma+1/2}(\sqrt{xu/4})}\right]\frac{K_{\gamma-1/2}(\sqrt{xu/4})}{K_{\gamma+1/2}(\sqrt{xu/4})}-\sqrt{\frac{xu}{4}}\right)\right].\label{lim-pol-odd.a}
\end{eqnarray}
Here we have used the rescaled function~\eref{kernel-F} which is one of the polynomials for $k=2m+1$ odd. The shift in the order of the polynomial from $l-2$ to $xp$, cf. Eqs.~\eref{kernel-F} and \eref{lim-pol-even}, has no effect on the limit since it can be absorbed in the limit $p\to\infty$. Furthermore we used the recurrence relation between the polynomials of even order and the polynomials of odd order, see Eq.~\eref{eq:polynomialsodddegree}, to derive the limit for the latter.
The ambiguous function $\widehat{c}_{px/2}(u/(4p))$ is chosen such that its limit exists and converges to $C(xu)$. Indeed this function only depends on the combination $xu$ because a scaling in $x$ yields an inverse scaling in $u$ by absorbing the scaling factor in the $p$ limit. We underline that also the constant $C(xu)$ is ambiguous and we may choose it in such a way that the prefactor in front of the Bessel function $I_{2\gamma+a+1}(\sqrt{xu})$ vanishes, i.e.
\begin{eqnarray}
 \fl&&\left.\lim\limits_{p\to\infty}\frac{(-1)^{xp-a}}{\Gamma[xp+1]}\left(\frac{u}{4p}\right)^{2\gamma+a}\partial_\kappa^{a}R_{xp+1}^{(\gamma)}\left(\kappa,\frac{u}{4p}\right)\right|_{\kappa=-u/(4p)}=\left(\frac{xu}{4}\right)^{(2\gamma+a)/2}I_{2\gamma+a}(\sqrt{xu})\nonumber\\ 
 \fl&&\times\left[\sqrt{\frac{xu}{4}}\left(\frac{K_{\gamma+1/2}(\sqrt{xu/4})}{K_{\gamma-1/2}(\sqrt{xu/4})}-\frac{K_{\gamma-1/2}(\sqrt{xu/4})}{K_{\gamma+1/2}(\sqrt{xu/4})}\right)-2\gamma-1-2a\right].\label{lim-pol-odd.b}
\end{eqnarray}
This expression is more compact than the one in Eq.~\eref{lim-pol-odd.a} such that we stick with this intermediate result (which is again simpler than in \cite{AGKWWprl}).

The normalization constants of the polynomials, see Eq.~\eref{norm-pol}, have the limit
\begin{equation}
\label{eq:MicroscopicLimitScalarProduct}
 \lim\limits_{p\to\infty} \frac{1}{\Gamma^2[xp+1](xp)^{2\gamma}}r_{xp/2}^{(\gamma)}\left(\frac{u}{4p}\right) =4.
\end{equation}
Moreover we need to express the limit of the sum of $p$ terms in order to deal with the two-point kernel~\eref{eq:Pfaffiankernel}. Let us consider a function $f_j$ where the limit $\lim_{p\to\infty}f_{px}=f(x)$ exists for all $x\in[0,1]$. Then the sum becomes
\begin{equation}
\label{eq:SumToIntegral}
\lim\limits_{p\to\infty} \frac{1}{p}\sum_{j=0}^{\lfloor(p+k-1)/2\rfloor}f_j = \frac{1}{2}\int\limits_0^1\dd{x} f(x).
\end{equation}
Combining this limit with Eqs.~\eref{lim-pol-even}, \eref{lim-pol-odd.b}, and \eref{eq:MicroscopicLimitScalarProduct} we find the asymptotics of the kernel
\begin{eqnarray}\label{asymp-kernel}
 \fl\Xi_{ab}^{(\gamma,\infty)}(u)&=&\lim\limits_{p\to\infty} \Xi_{ab}^{(\gamma,p)}\left(\kappa,\frac{u}{4p}\right)\\
 \fl&=&\left.\lim\limits_{p\to\infty}(-1)^{a+b}\left(\frac{u}{4p}\right)^{2\gamma+a+b+1}\partial_{\kappa_1}^a\partial_{\kappa_2}^b\mathcal{K}_{p}\left(\kappa_1,\kappa_2,\frac{u}{4p}\right)\right|_{\kappa_1=\kappa_2=-u/(4p)}\nonumber\\
 \fl&=&\frac{1}{4}\int\limits_0^{\sqrt{u}/2} \dd x x^{a+b+1}\biggl[2(b-a)I_{2\gamma+a}(2x)I_{2\gamma+b}(2x)\nonumber\\
 \fl&&+\frac{K_{\gamma-1/2}(x)}{K_{\gamma+1/2}(x)}\left[x\left(\frac{K_{\gamma+1/2}(x)}{K_{\gamma-1/2}(x)}-\frac{K_{\gamma-1/2}(x)}{K_{\gamma+1/2}(x)}\right)-2\gamma-1\right]\nonumber\\
 \fl&&\times\left(I_{2\gamma+a}(2x)I_{2\gamma+b+1}(2x)-I_{2\gamma+a+1}(2x)I_{2\gamma+b}(2x)\right)\nonumber\\
 \fl&&+\frac{K_{\gamma-1/2}(x)}{K_{\gamma+1/2}(x)}\left(2bI_{2\gamma+a+1}(2x)I_{2\gamma+b}(2x)-2a I_{2\gamma+a}(2x)I_{2\gamma+b+1}(2x)\right)\biggl].\nonumber
\end{eqnarray}
Recall that the asymptotics for even and odd $p$ yields the same answer, and so we have obtained this limit by choosing $p=2L$ even.

When considering the particular cases of the gap probability ($\gamma=0$) and the distribution of the smallest eigenvalue ($\gamma=1$) it is more enlightening to express the Bessel functions involved in terms of more explicit functions,
\begin{eqnarray}
 K_{1/2}(z)&=&K_{-1/2}(z)=\sqrt{\frac{\pi}{2}}\frac{1}{\sqrt{z}}e^{-z},\label{Bessel-expl.a}\\
 K_{3/2}(z)&=&\left(1+\frac{1}{z}\right)K_{1/2}(z)=\sqrt{\frac{\pi}{2}}\left(\frac{1}{z^{1/2}}+\frac{1}{z^{3/2}}\right)e^{-z}.\label{Bessel-expl.b}
\end{eqnarray}
Then the gap probability reads
\begin{eqnarray}
  \fl \mathcal E_{2k}(u) &=&\left(\prod\limits_{l=0}^{k-1}\frac{4^{l+1}(2l)!}{l!}\right)u^{-k^2/2}e^{-u/8-\sqrt{u/4}}\nonumber\\
  \fl&&\times\left\{\begin{array}{cl} \displaystyle\Pf_{0\leq a,b \leq k-1}\left[  \Xi_{ab}^{(0,\infty)}(u)\right], & k\in2\mathbb{N}_0,\\
  \displaystyle \frac{\sqrt{u}}{4}\Pf_{0\leq a,b \leq k-1}\left[
\begin{array}{ll}
\Xi_{ab}^{(0,\infty)}(u) & \xi_{a}^{(0,\infty)}(u)\\
-\xi_{b}^{(0,\infty)}(u) &0\\
\end{array}\right], & k\in2\mathbb{N}_0+1
  \end{array}\right.
  \label{Gap-asy}
\end{eqnarray}
with
\begin{eqnarray}
 \fl\xi_a^{(0,\infty)}\left(u\right)&=&\left(\frac{u}{4}\right)^{a/2}\left[I_{a}(\sqrt{u})+I_{a+1}(\sqrt{u})\right],
\label{Gap-asy-a}\\
\fl\Xi_{ab}^{(0,\infty)}(u)&=&\frac{1}{4}\int\limits_0^{\sqrt{u}/2} \dd x x^{a+b+1}\biggl[2(b-a)I_{a}(2x)I_{b}(2x)
\label{Gap-asy-b}\\
 \fl&&+(2b+1)I_{a+1}(2x)I_{b}(2x)-(2a+1) I_{a}(2x)I_{b+1}(2x)\biggl].\nonumber
\end{eqnarray}
The results in Ref.~\cite{AGKWWprl}, have three typos. First, the factor $x^{a+b+1}$ ($z^{(a+b)/2}$ in the notation therein) is missing in the integral for the two-point kernel see Eqs.~(25) therein. Second, the indices of the Pfaffian in the case of odd $k$ should go from $0$ to $k-1$, see Eqs.~(27) and (29) therein.

The gap probability seems to diverge at $u\to0$ due to the terms $u^{-k^2/2}$ and $u^{-(k^2-1)/2}$. However the kernels vanish as $\Xi_{ab}^{(0,\infty)}(u)\propto u^{a+b+1}$ and $\xi_{a}^{(0,\infty)}(u)\propto u^{a}$ for $u\ll 1$ such that both terms cancel and the gap probability behaves as a constant, especially it is normalized to $\mathcal E_{2k}(0)=1$.
%We use
%\begin{equation}
%\eqalign{
%\fl\quad&\int\limits_0^1\dd{x} y^{c} \text{I}_{\mu}\left(y\sqrt{u}\right)\text{I}_{\nu}\left(y\sqrt{u}\right) = 2^{-\sigma-1} u^{\sigma} \Gamma (\sigma+1) \Gamma \left(\frac{1}{2} (c+\sigma+1)\right) \, \\\fl\quad&_3\tilde{F}_4\left(\frac{1}{2} (\sigma+1),\frac{1}{2} (\sigma+2),\frac{1}{2} (c+\sigma+1);\mu+1,\frac{1}{2} (c+\sigma+3),\nu+1,\sigma+1;u\right)~,}
%\end{equation}
%where $\sigma=\mu+\nu$ and $ _3\tilde{F}_4$ is the regularized counterpart of $_3F_4$, c.f. \Ref{AbramowitzStegun}.

\begin{figure}
 \centerline{\includegraphics[width=0.7\textwidth]{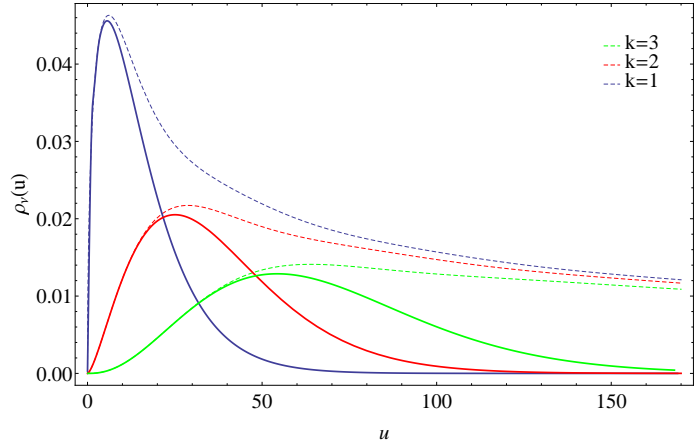}}
 % Egap.png: 1199x790 pixel, 72dpi, 42.29x27.87 cm, bb=0 0 1199 790
 \caption{The distribution of the smallest eigenvalue~\eref{dis-asy} (solid curves) smoothly fits with the microscopic level density~\eref{eq:intro:density} (dotted curves) around the origin. Only for larger values of $u$ deviations due to the contribution of the second and larger eigenvalues become visible.
A similar plot can be made for $\nu=0$ which we have omitted for a better readability of the figure.}
 \label{fig:dist}
 \end{figure}

The distribution of the first eigenvalue is
\begin{eqnarray}
  \fl \mathcal P_{2k}(u)&=&\frac{1}{8}\left(\prod\limits_{l=0}^{k-1}\frac{4^{l+1}(2l)!}{l!}\right)(\sqrt{u}+2)u^{-(k^2+1)/2}e^{-u/8-\sqrt{u/4}}\nonumber\\
  \fl&&\times\left\{\begin{array}{cl} \displaystyle\Pf_{0\leq a,b \leq k-1}\left[  \Xi_{ab}^{(1,\infty)}(u)\right], & k\in2\mathbb{N}_0,\\
  \displaystyle \frac{1}{\sqrt{u}}\Pf_{0\leq a,b \leq k-1}\left[
\begin{array}{ll}
\Xi_{ab}^{(1,\infty)}(u) & \xi_{a}^{(1,\infty)}(u)\\
-\xi_{b}^{(1,\infty)}(u) &0\\
\end{array}\right], & k\in2\mathbb{N}_0+1.
  \end{array}\right.
\label{dis-asy}
\end{eqnarray}
The behaviour of the distribution at the origin can be read off from the kernels which are
\begin{eqnarray}
 \fl\xi_a^{(1,\infty)}\left(u\right)&=&\left(\frac{u}{4}\right)^{a/2+1}\left[I_{a+2}(\sqrt{u})+\frac{\sqrt{u}}{\sqrt{u}+2}I_{a+3}(\sqrt{u})\right],
\label{dis-asy-a}\\
\fl\Xi_{ab}^{(1,\infty)}(u)&=&\frac{1}{4}\int\limits_0^{\sqrt{u}/2} \dd x x^{a+b+1}\biggl[2(b-a)I_{a+2}(2x)I_{b+2}(2x)
\label{dis-asy-b}\\
 \fl&&\hspace*{-1.5cm}+\frac{x}{x+1}\left(\left(2b+\frac{x+2}{x+1}\right)I_{a+3}(2x)I_{b+2}(2x)-\left(2a+\frac{x+2}{x+1}\right) I_{a+2}(2x)I_{b+3}(2x)\right)\biggl].\nonumber
\end{eqnarray}
Since  $\Xi_{ab}^{(1,\infty)}\left(u\right)\propto u^{a+b+3}$ and $\xi_a^{(1,\infty)}\left(u\right)\propto u^{a+2}$ for $|u|\ll1$ we have $\mathcal P_{2k}(u)\propto u^{k-1/2}$ agreeing with the behaviour of the microscopic level density~\eref{eq:intro:density}. We remark that the term $J_\nu(\sqrt{u})/\sqrt{16u}\sim u^{(\nu-1)/2}$ in the expression~\eref{eq:intro:density} is the dominant term in the limit $u\to0$. The same behaviour of both distributions around the origin is inherent because the level density is governed by the smallest eigenvalue in this regime. Only for larger argument $u$ the other eigenvalues start to contribute to the level density, cf. Fig.~\ref{fig:dist}.

\begin{figure}
 \centerline{\includegraphics[width=0.7\textwidth]{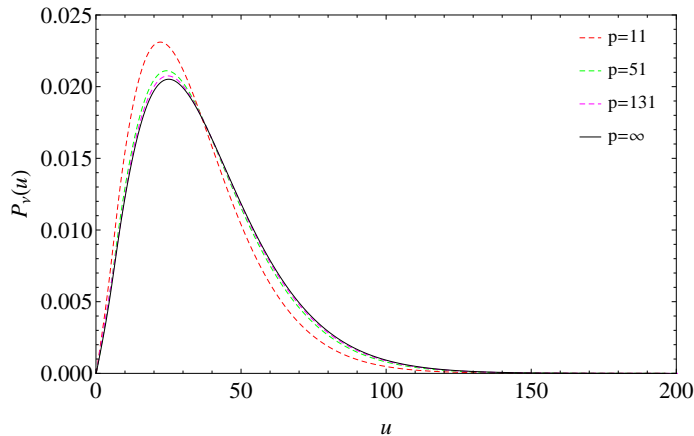}}
 % Egap.png: 1199x790 pixel, 72dpi, 42.29x27.87 cm, bb=0 0 1199 790
 \caption{Illustration of the rate of convergence by comparing the finite $p$ result~\eref{eq:smallDist:FinalOddpEvenAlpha} for $p=11,51,131$ with the universal limit $p\to\infty$ ~\eref{dis-asy} for the distribution of the smallest eigenvalue. The index of the Wishart matrix is chosen $\nu=2k=4$ as an example.}
 \label{fig:comp}
 \end{figure}

The results~\eref{Gap-asy} and \eref{dis-asy} are our third main result. We emphasize that these results do not only describe the smallest eigenvalue of an artificial system, namely real chiral Gaussian random matrices with an even index $\nu=2k$, but also of physical systems. Due to universality not only the level density~\eref{eq:intro:density} has to agree with those from physical systems like QCD or mesoscopic systems. Also the distributions of the smallest eigenvalues
of those physical systems  have to follow the same distributions of random matrix theory in the limit of its applicability, e.g. see  \cite{JacTilo,carlo,handbook} and references therein. This agreement should already happen at moderate system sizes. In Fig.~\ref{fig:comp} we compare the analytic results~\eref{eq:smallDist:FinalOddpEvenAlpha} and \eref{eq:smallDist:FinalOddpOddAlpha} of the distribution of the smallest eigenvalue at finite $p$ with the microscopic limit~\eref{dis-asy}. This comparison underlines how fast the convergence to the universal result happens. Although this comparison is done for random matrices which have particular advantages in comparison to real systems we expect that also physical systems should display a rapid convergence to the universal result. 

%The reason is that the first eigenvalue lies in the middle of the bulk of the spectrum. This eigenvalue is best screened from finite volume effects.

%%%%%%%%%%%%%%%%%%%%%%%%%%%%%%%%%%%%%%%%%%%%%%%%%%%%%%%%%%%%%%%%%%%%%%%

\section{The Correlated Wishart-Laguerre Ensemble}\label{sec:numericalsimulations}

We are now interested in the effects of a fixed, non-trivial correlation matrix $C$ not proportional to the identity matrix on the distribution of the smallest eigenvalue, see Eq.~\eref{eq:intro:wishartP}. Such a correlation matrix can naturally encode system specific information. In time series analysis such a correlation may encode correlations between companies in finance \cite{S12}, seasonal effects in climate research \cite{RAP10} or organized crime in criminal defence~\cite{TEP11}. However usually such correlation matrices have no microscopic limit in time series. But also in QCD and mesoscopic systems which  exhibit such a microscopic limit correlations may appear. These correlations encode the structure of space-time and the choice of the gauge theory for example. They are system specific informations and may have an influence on the smallest eigenvalues.

We expect that the correlation matrix has no influence on the smallest eigenvalues as long as its eigenvalues have a finite distance to the origin. Then the screening of the infinitely many eigenvalues between the smallest eigenvalue of $W$ and the eigenvalues of $C$ is strong enough. This was also shown in \cite{WG}. For this purpose we choose a non-trivial empirical correlation matrix $C\neq \Id_p$. In Fig.~\ref{fig:comp} we compare  Monte-Carlo simulations with such an empirical correlation matrix $C$ and the universal result~\eref{dis-asy}.  The matrix size is chosen such that $(n-p)/p=\nu/p\ll 1$ where $p=200$ and $\nu=2k=0,2,4$. The perfect agreement underlines that correlations in the Wishart matrix have a very weak effect on the spectral statistics of the smallest eigenvalue.

\begin{figure}[t]
 \centering
 \includegraphics[width=0.7\textwidth]{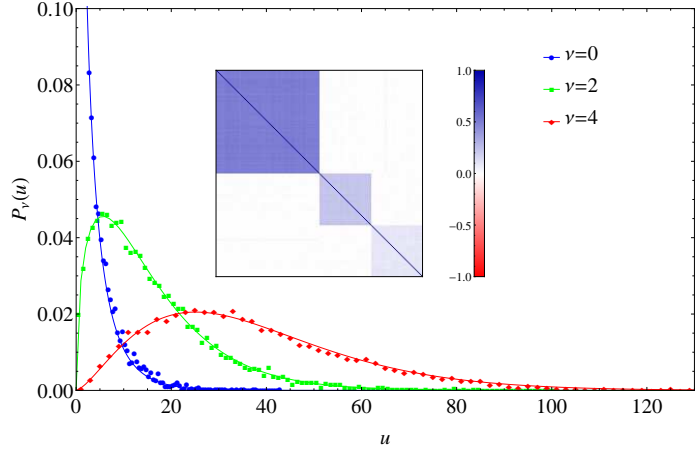}
 % Pmin.png: 1216x777 pixel, 72dpi, 42.89x27.41 cm, bb=0 0 1216 777
 \caption{Comparison of Monte Carlo simulations (symbols) with a non-trivial empirical correlation matrix $C$ at finite $p=200$, 
and the microscopic limit $p\to\infty$ of the distribution of the smallest eigenvalue~\eref{dis-asy} (solid lines). The empirical correlation matrix is shown in the inset. We have generated $10000$ correlated Wishart random matrices of sizes $200\times(200+\nu)$ with $\nu=2k=0,2,4$.}
 \label{fig:pmincor}
\end{figure}

Again we emphasize that we have not looked at the situation where $C$ develops a spectrum where some eigenvalues lie on the scale $1/p$. Nor have we looked at the situation of doubly correlated Wishart-Laguerre ensembles. However for the latter kind of ensembles we expect a similar if not exactly the same behaviour like one-sided correlated Wishart-Laguerre ensembles as considered here.

%%%%%%%%%%%%%%%%%%%%%%%%%%%%%%%%%%%%%%%%%%%%%%%%%%%%%%%%%%%%%%%%%%
\section{Conclusion}\label{sec:conclusion}

We have addressed and solved an open problem in the real Wishart-Laguerre ensemble also known as the chiral Gaussian orthogonal ensemble of rectangular $p\times n$ random matrices. We computed the distribution of the smallest eigenvalue and its integral, the gap probability that the vicinity of the origin is empty of eigenvalues. 
To this aim we have established that an integrable Pfaffian structure holds also when $p-n=\nu$ measuring the rectangularity (or topology in the field theory application) is even. Such a Pfaffian structure was previously only known when $\nu=2k+1$ is odd. So far for an even rectangularity a recursive construction in $p$ led to closed form expressions for $\nu=0,2$ only.  
%A recursive construction in $p$ that gave closed form expressions only for $\nu=0,1,2,3$. In this recursion such structure was obscured. 
In view of the various applications it would be unnatural to restrict oneself to odd $\nu$, and not to expect for such an integrable structure to exist. 
However, the recursive construction (and closed results for $\nu=0,2$) already revealed the appearance of special functions for finite $p$ which are absent for $\nu$ odd, namely Tricomi's confluent hypergeometric functions. From our construction we now better understand why they appear through the expectation value of the square root of characteristic polynomials which are among the building blocks for the quantities in question. 

On a technical level our computation was possible due to the combination of the method of skew-orthogonal polynomials, though with a non standard weight, and the map of our building blocks onto invariant co-set integrals derived by bosonisation. In an initial step the computation of the the gap probability and the distribution of the smallest eigenvalue requires the evaluation of expectation values of characteristic polynomials raised to integer/half-integer powers for $\nu$ odd/even, respectively. Hence the problem exhibits an increased level of difficulty in computing these objects for even $\nu$. By including the square root for even $\nu$ into the weight function we were back to expectation values of integer powers, which are known to be expressible through Pfaffian determinants of kernels and skew-orthogonal polynomials. The price we had to pay was to compute the latter for a non-standard weight including the square root. This was done by expressing the polynomials and kernel themselves through 
expectation values, mapping these back to matrix integrals and computing them via bosonisation.

Indeed one can also consider the distribution and the cumulative distribution of the second to smallest eigenvalue, third to smallest eigenvalue etc. These quantities can be simply deduced from our results, too, because we calculated a quite explicit expression of the kernel at finite $p$. In particular we found a Christoffel-Darboux-like formula which expresses the sum over $(p+k)$ terms in a sum over twelve terms, only.

The Pfaffian structure enabled us to take the microscopic large-$p$ limit at the origin, while keeping $\nu=2k$ fixed. In this limit we could show that the distinction between even and odd $p$ for finite $p$ becomes immaterial. We found results in terms of a Pfaffian comprising the limiting kernel for even $k$, plus an additional column and row for $k$ odd, both for the gap probability and the smallest eigenvalue.

Our results are universal for non-Gaussian potentials, as inherited from the universality of the known density correlation functions. We have checked that our findings follow the microscopic density for small argument, and that our finite-$p$ results, which we have confirmed through numerical simulations, converge towards the universal limit. Furthermore, we have also studied numerically the distribution of the smallest eigenvalue in an example of a correlated Wishart-Laguerre ensemble. We found that for moderate $p$ it already follows the universal limiting distribution for several values of $\nu$.

The computation of products of ratios of characteristic polynomials that also include square roots is in general an open question in random matrix theory. The structure of the results we obtained on a subset of such correlators should be relatively easy to translate to the Gaussian orthogonal ensemble, where such correlation functions enjoy further applications, e.g. in Quantum Chaos.
It is very plausible that our universal result will also apply when introducing a fixed trace constraint, as it is known for odd $\nu$. What is less clear is whether a corresponding representation in terms of hypergeometric functions of matrix arguments exist, having the advantage that they can be continued to real $\beta>0$.
It would also be very interesting to see, if and when the universality at the origin breaks down for the correlated Wishart-Laguerre ensemble.
A further open question is the computation of the gap probability and smallest eigenvalue distribution in the chiral Gaussian symplectic ensemble with $\beta=4$. Apart from $\nu=0$ only Taylor expansions exist so far. However, following similar ideas as in the present work, Pfaffian structures for these quantities exist and should be universal. Work in this direction is currently under way.

\section*{Acknowledgements}
We kindly acknowledge support from the German Research Council (DFG) via the Sonderforschungsbereich Transregio 12, ``Symmetries and Universality in Mesoscopic Systems'' (T.W., T.G. and G.A.) and the Alexander von Humboldt-Foundation (M.K.).

\end{document}